\documentclass[aip,jrse,reprint,onecolumn]{revtex4-1} 
\usepackage{graphicx}
\usepackage{dcolumn}
\usepackage{bm}
\usepackage{amssymb}
\usepackage{amsmath}
\usepackage{amsfonts}
\usepackage{mathtools}
\usepackage{comment}
\usepackage{lineno}
\usepackage{soul}
\usepackage{natbib}
\usepackage{color}      
\usepackage{soul}
\usepackage[usenames,dvipsnames,svgnames,table]{xcolor} 
\usepackage[colorlinks=true,
            linkcolor=blue,
            urlcolor=blue,
            citecolor=BrickRed]{hyperref}
            
\usepackage{mathrsfs}  
\usepackage{longtable}
\usepackage{tabulary}
\newcommand{\blue}{\color{black}} 

\newcommand{\black}{\color{black}}

\usepackage{mathtools}
\usepackage{amssymb}
\usepackage{amsmath}
\usepackage[abs]{overpic}
\usepackage{xcolor,varwidth}            

\begin{document}
\title{Data-driven fluid mechanics of wind farms: A review}
\author{Navid Zehtabiyan-Rezaie}
\affiliation{Department of Mechanical and Production Engineering, Aarhus University, 8000 Aarhus C, Denmark}%
\author{Alexandros Iosifidis}
\affiliation{Department of Electrical and Computer Engineering, Aarhus University, 8000 Aarhus C, Denmark}
\affiliation{Center for Digitalization, Big Data, and Data Analytics, Aarhus University, 8000 Aarhus C, Denmark}
\author{Mahdi Abkar}
\email{abkar@mpe.au.dk}
\affiliation{Department of Mechanical and Production Engineering, Aarhus University, 8000 Aarhus C, Denmark}
\affiliation{Center for Digitalization, Big Data, and Data Analytics, Aarhus University, 8000 Aarhus C, Denmark}

\begin{abstract}
With the growing number of wind farms over the last decades and the availability of large datasets, research in wind-farm flow modeling – one of the key components in optimizing the design and  operation of wind farms – is shifting towards data-driven techniques. However, given that most current data-driven algorithms have been developed for canonical problems, the enormous complexity of fluid flows in real wind farms poses unique challenges for data-driven flow modeling. These include the high-dimensional multiscale nature of turbulence at high Reynolds numbers, geophysical and atmospheric effects, wake-flow development, and incorporating wind-turbine characteristics and wind-farm layouts, among others. 
In addition, data-driven wind-farm flow models should ideally be interpretable and have some degree of generalizability. The former is important to avoid a lack of trust in the models with end-users, while the most popular strategy for the latter is to incorporate known physics into the models.  
This article reviews a collection of recent studies on wind-farm flow modeling covering both purely data-driven and physics-guided approaches. We provide a thorough analysis of their modeling approach, objective, and methodology, and specifically focus on the data utilized in the reviewed works.
\end{abstract}

\maketitle


\section{Introduction} 
\label{section:Introduction}
\begin{figure}
	\centering
	\includegraphics[width=0.75\textwidth]{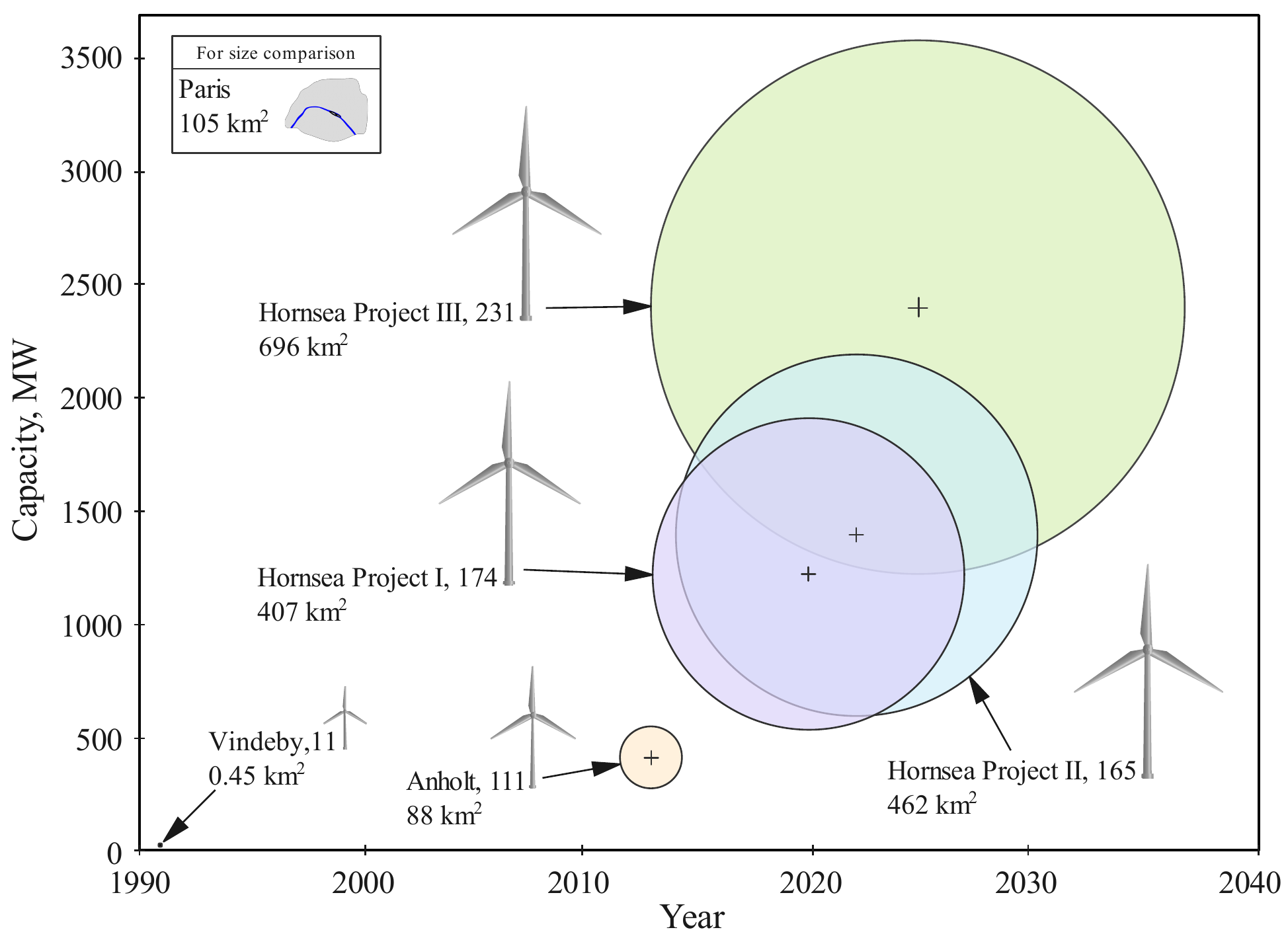}
	\caption{The change in size and capacity of several offshore wind farms over time. \blue The numbers next to the turbines indicate the number of wind turbines in the corresponding wind farm. \black Circle centers indicate the capacity of wind farms and the year of operation starting. Circles' area show the area covered by each wind farm.}
	\label{fig:Fig1}
\end{figure}

The increasing demand for electric power finds no answer other than clean energy when the environmental concerns related to climate change are on the table. 
The policies of net-zero greenhouse emission followed by many countries \cite{Potrc2021,Zappa2019} raise the hope that clean energy can produce a significant fraction of the world's electric power needs in the future. But to reach net-zero emission by 2050, the annual wind-energy installation must increase dramatically \cite{GLOBALWINDREPORT2021}.
Wind-power technologies are mature and wind energy already has a share in the electric power production of many countries.
\blue During the past decades, tens to hundreds of large-scale turbines have been put together in wind farms. As a result, studying the complex interplay between wind turbines and the atmospheric boundary layer (ABL) has become an important and challenging topic in wind-energy research\cite{veers2019grand}, and now the concern is extended to wake interactions between multiple wind farms \cite{platis2018first,lundquist2019costs,schneemann2020cluster}. \black
Fig.~\ref{fig:Fig1} schematically shows the trend of increasing size and complexity in some examples from the offshore wind industry starting with the world's first offshore wind farm in 1991 and continuing to grow to large-scale offshore wind-farm clusters. 
The prerequisites of this shift are modeling, simulation, and optimization of wind-energy systems from individual-turbine level to multiple-wind-farm level.

\blue The objectives of many fluid-mechanics-related studies in wind energy include wind-turbine aerodynamics, wind-farm flow modeling, and wind-farm flow control.
The focus of wind-turbine aerodynamic analysis is mainly on the interactions between the blades and unsteady flow to optimize the rotor aerodynamic performance (see the review of Refs. \cite{hansen2006state,hansen2011review,wang2012brief}). 
Wind-farm flow control often seeks to optimize power production, structural loads, and operation costs using an individual-turbine-level approach through wake steering and downregulation (see the review of  Refs. \cite{kheirabadi2019quantitative,Shapiro2021,nash2021wind,houck2022review}).
Wind-farm flow modeling includes analysis of fluid flow in wind farms, evaluation of turbines' performance and, consequently, wind-farm power generation. \black 
The first group of studies focuses on the fluid mechanics of wind farms, two-way interactions between wind turbines and the ABL,  thermal stability effects, and wake recovery. The second group of studies put their focus on turbine- or farm-level power-output evaluation. 
\blue
The studies on wind-farm flow modeling utilize either physics-driven or data-driven modeling approaches. An overview of wind-farm flow modeling with the focus on physics-driven methods can be found in Refs. \cite{Vermeer2003,Sanderse2011,Stevens2017,PorteAgel2019Review}. 
This paper presents a review of the recent studies utilizing data-driven techniques for wind-farm flow modeling.
\black

Due to the presence of the non-stationary phenomena in wind farms, e.g. wake meandering, dynamic wake modeling and power prediction may be necessary to monitor and optimize the real-time operation of wind farms. Therefore, we have also attempted to identify the objective of the reviewed works regarding static (steady-state) and dynamic (transient) modeling of wind farms, including static/dynamic wake modeling and static/dynamic power prediction. 
In this paper, we also take a closer look at the data utilized in the reviewed works, as the primary element of data-driven modeling. We aim to clarify the data type, source, and availability for the readers. A thorough survey of the data utilized in the reviewed studies can also identify the necessary future steps to be taken in terms of generation and sharing of data.

\begin{table}[ht] 
\centering
\caption{List of abbreviations used in the present review article.}
\label{Table:Abb} 
\begin{tabular} {p{1.2cm} p{6.6cm} p{1.2cm} p{6cm}}
\hline
\small
ABL	&	Atmospheric boundary layer	&	LiDAR	&	Light detection and ranging	\\
ANN	&	Artificial neural network	&	LSTM	&	Long short-term memory	\\
BD	&	Blocking distance	&	ML	&	Machine learning	\\
BPNN	&	Backpropagation neural network	&	NIROM	&	Non-intrusive reduced-order model(ing)	\\
BR	&	Blockage ratio	&	ODE	&	Ordinary differential equation	\\
CFD	&	Computational fluid dynamics	&   	PCA	&	Principal component analysis	\\
CNN	&	Convolutional neural network	&	PDDM	&	Purely data-driven model(ing)	\\
DDM	&	Data-driven model(ing)	&	PDM	&	Physics-driven model(ing)	\\
DMD	&	Dynamic mode decomposition	&	PGDDM	&	Physics-guided data-driven model(ing)	\\
ERT	&	Extremely randomized trees	&	POD	&	Proper orthogonal decomposition	\\
FLORIS	&	Flow redirection and induction in steady-state	&	RANS	&	Reynolds-averaged Navier-Stokes	\\
GA	&	Genetic algorithm	&	RBF	&	Radial basis function	\\
GAM	&	Generalized additive model	&	RF	&	Random forest	\\
GAN	&	Generative adversarial network	&	RNN	&	Recurrent neural network	\\
GBM	&	Gradient boosting model	&	ROM	&	Reduced-order model(ing)	\\
GBR	&	Gradient boosting regression	&	SCADA	&	Supervisory control and data acquisition	\\
GNN	&	Graph neural network	&	SOWFA	&	Simulator for wind farm applications	\\
GP	&	Gaussian process	&	SVM	&	Support vector machine	\\
GRNN	&	General regression neural network	&	SVR	&	Support vector regression	\\
IROM	&	Intrusive reduced-order model(ing)	&	TI	&	Turbulence intensity	\\
LES	&	Large-eddy simulation	&	TKE	&	Turbulent kinetic energy	\\

\hline
\end{tabular}
\end{table}
\normalsize

\begin{figure}[ht] 
	\centering
	\includegraphics[width=1\textwidth]{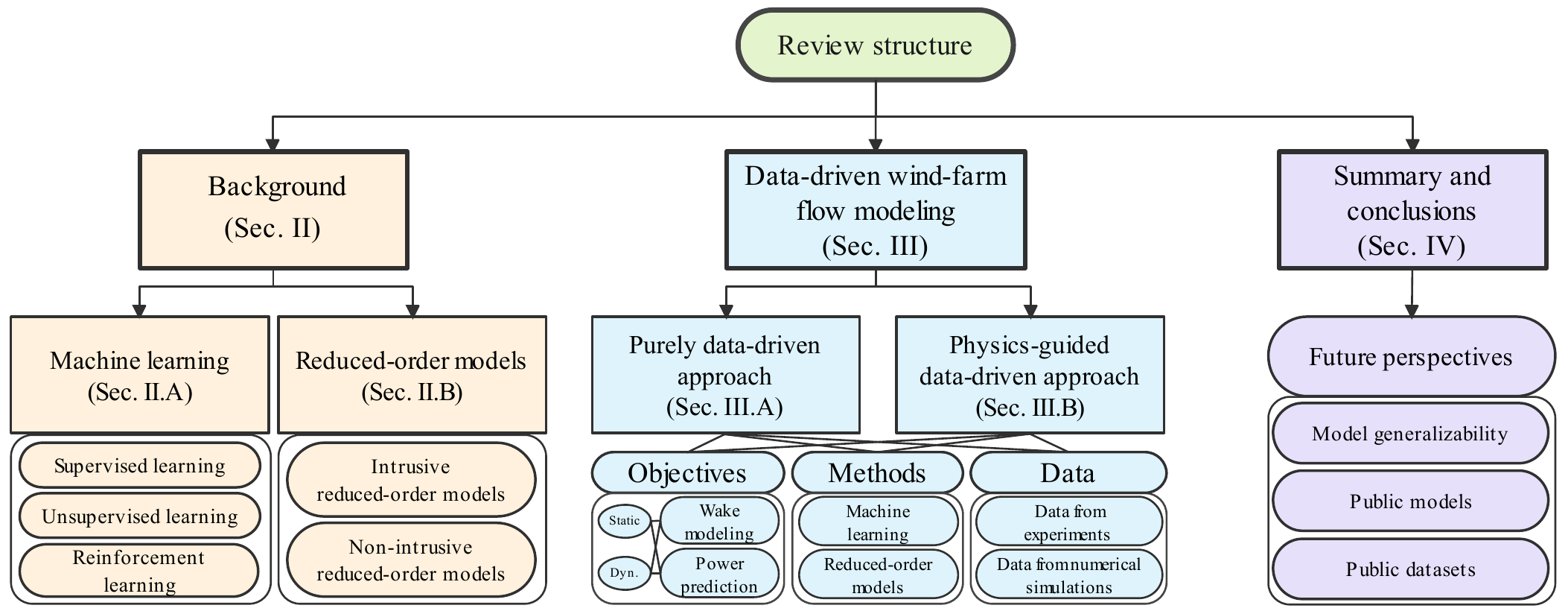}    
 	\caption{The structure of the present review article.}
	\label{fig:Fig2}
\end{figure}

This review paper is organized as follows: Section~\ref{section:Background} provides a short background on data-driven modeling (DDM) approaches and methods. In Section~\ref{section:Results}, the approach, objective, utilized data, and methodology of the selected papers are analyzed. Finally, in Section~\ref{section:Conclusions}, a summary and future perspectives are given.
Fig.~\ref{fig:Fig2} shows the structure of the review.
The abbreviations used in this paper are defined in Table \ref{Table:Abb}. 

\section{Background}
\label{section:Background}
\begin{figure}[htbp] %
	\centering
    \includegraphics[scale=0.4]{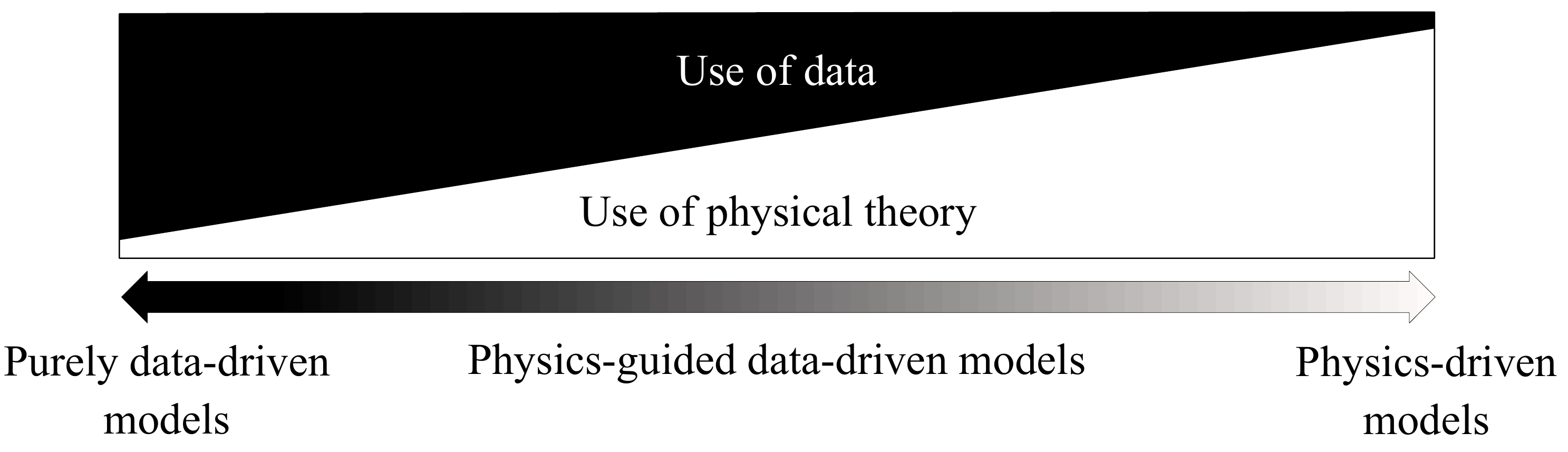}
	\caption{Overview of modeling approaches.}
	\label{fig:Fig3}
\end{figure}

Researchers have always been interested in modeling the flow in wind farms to obtain an assessment of wind-farm power production, as a function of several effective parameters.
This is necessary for the design, analysis, and control of wind farms. 
As shown in Fig.~\ref{fig:Fig3}, \blue purely data-driven models (PDDMs) \black and physics-driven models (PDMs) are the two ends of the models' spectrum. \blue PDDMs\black, also called black-box models, attempt to make a connection between inputs and outputs without paying much attention to the physics of the problem. On the other hand, PDMs, which follow a white-box approach, are based on extended mathematical equations that represent the governing physics of the problem, for example, the Navier-Stokes equations consist of the conservation of mass, momentum, and energy equations. 

A simpler example of PDMs is the analytical wake models, which are used to predict wind-turbine or wind-farm flows by considering several simplifying assumptions. The analytical wake models based on either or both mass and momentum conservation laws can estimate the wake growth behind a single wind turbine (see the review of Refs. \cite{goccmen2016wind,Archer2018,PorteAgel2019Review} and references therein) and evaluate the interaction of wakes of several wind turbines through superposition techniques \cite{lissaman1979energy,katic1986simple,voutsinas1990analysis,Niayifar2016,Zong2020,Bastankhah2021,lanzilao2022new}. 
These models can be used for wind-farm performance evaluation and layout optimization due to their low computational cost \cite{chowdhury2012unrestricted,shakoor2016wake,tao2020nonuniform}. The computational fluid dynamics (CFD) techniques including Reynolds-averaged Navier-Stokes (RANS) equations and large-eddy simulation (LES) can be listed as the mid- and high-fidelity methods in wind-energy applications. They force high computational costs compared to the analytical models but can capture the complex flow in the wind farms (see, e.g., Refs. \cite{Vermeer2003,Sanderse2011,Meneveau2019}). In some recent work \cite{iungo2015data,king2018data,steiner2021classifying,Steiner2022}, data-driven methods have been utilized to improve RANS simulations of wake flow in wind farms by using LES data, but since they are under the general category of data-driven RANS modeling, are not covered in this review.  

Some physical phenomena can be accurately described by first-principle models, but direct solving of the PDMs often imposes high computational costs. In this case, DDMs can fill the void. Today, DDMs are playing a vital role when the simplified PDMs cannot accurately describe the system's behavior in real operating conditions \cite{montans2019data}. 
To more accurately predict the actual performance of systems, given the current computing power and data availability, we can combine \blue purely \black data-driven and physics-driven approaches resulting in a gray-box approach named physics-guided data-driven models (PGDDM) \cite{Legaard2021}. In this review article, \blue PDDM and PGDDM \black approaches used for wind-farm flow modeling by utilization of machine learning (ML) tools and reduced-order models (ROM) are covered.  

\subsection{Machine learning}
\begin{figure}[ht] 
	\centering
	\includegraphics[width=1\textwidth]{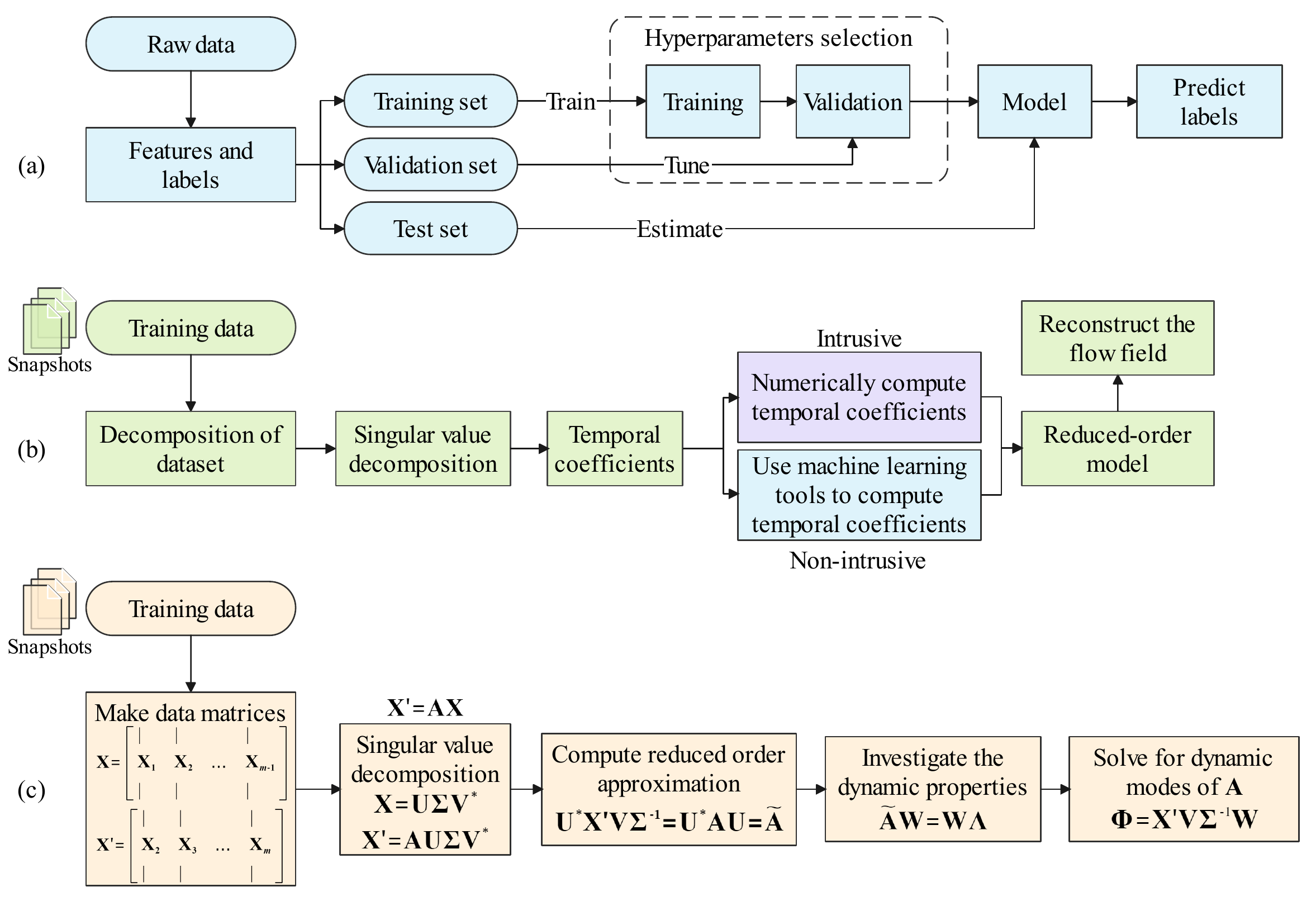}
	\caption{Algorithm of data-driven models: (a) ML (b) IROM and NIROM (c) DMD.}
	\label{fig:Fig4}
\end{figure}

As shown in Fig.~\ref{fig:Fig4}a, the objective of most ML algorithms is to learn a mapping from input data to output data for the purpose of prediction or classification. Clustering methods, as well as methods for determining very important data features to reduce dimensions, also belong to this category \cite{Goodfellow2016}.
Regardless of the application and purpose of the algorithm, ML approaches can be divided into three categories based on the learning paradigm they follow, i.e. those following supervised, unsupervised, or reinforcement learning. Supervised learning is used to learn hidden structures where output data is  known. Popular methods of supervised learning include artificial neural networks (ANN) \cite{nielsen2015neural}, decision trees \cite{loh2011classification}, k-nearest neighbors \cite{peterson2009k}, linear and polynomial regressions \cite{fahrmeir2013regression}, and support vector machines (SVM) \cite{scholkopf2002learning}.

\blue As ANN is one of the most widely used ML tools, it is briefly described here. For a given input, an ANN provides an output which takes the form of either a regression mapping or a probability-like response. This output is calculated by performing a series of transformations of the input, corresponding to the layers in the ANN topology. Determining these data transformations usually involves a training phase during which the network's weights and bias values are modified in an end-to-end manner for mapping the training data to known targets using a gradient-based iterative optimization process called backpropagation \cite{zaras2022DLbook}. Depending on the number of layers in an ANN topology, the ANN is a shallow network or a deep learning model. The type of transformations performed by the network layers depends on the type of neurons forming them. For example, multilayer perceptrons are formed by neurons performing linear transformations, convolutional neural networks (CNN) contain neurons performing convolution transformation of their inputs \cite{raitoharju2022DLbook}, recurrent neural networks (RNN) contain neurons that can aggregate information over a time-series input \cite{tsantekidis2022DLbook}, and graph convolutional networks aggregate information over multiple inputs the connections of which are indicated by a graph structure \cite{heidari2022DLbook}. While it is a common practice to use the same type of transformation for all hidden layer neurons, ANNs formed by heterogeneous set of neurons and nonlinear transformations have also been proposed \cite{kiranyaz2017pop,tran2020heterogeneous,kiranyaz2020Operational,kiranyaz2021selfonn}. After performing the transformation of its input, each neuron is equipped with a nonlinear activation function, thus the transformation performed by a neural layer is nonlinear. Deep learning models can handle more complex tasks with nonlinear behavior than shallower ANNs and other ML models, but they usually require large amounts of data for training. ANNs have a high degree of flexibility due to their end-to-end training process, but interpretation of their output is challenging \cite{Goodfellow2016}. \black

Unlike supervised learning, the output data is not known in unsupervised learning. Clustering is one of the most common problems in unsupervised learning \cite{jain1999data}. Another application of unsupervised learning is dimensionality reduction. The principal component analysis (PCA) method  \cite{jolliffe2016principal} \blue is an unsupervised learning method which \black identifies the principal axes of the data by examining the variations among the input data. The linear PCA is known as proper orthogonal decomposition (POD) \cite{chatterjee2000introduction} among researchers working on ROMs.
\blue
PCA aims to project an $n$-dimensional data set onto a $k$-dimensional surface, so that the reconstruction error is minimized. In other words, PCA finds a set of $k$ vectors and projects the $n$-dimensional data onto the linear subspace spanned by those vectors \cite{Ringner2008}. Mapping the projected data to the original space then leads to the minimum error a linear method can achieve in terms of $l_2$-norm. For this reason, PCA has been widely used for dimensionality reduction and compression, i.e., for reducing the required disk space and memory for storing the data, and also for speeding-up a learning algorithm. PCA sometimes discovers underlying hidden structures in data \cite{tenenbaum2000global}. It is worth mentioning that PCA can be reshaped into an ANN called autoencoder which is formed by an input layer, a hidden layer (bottleneck), and an output layer, where the data representations in the hidden layer correspond to the projected data representations. Such an ANN is trained to reconstruct in its output the data introduced in its input. Extensions of this type of ANN to include nonlinear activation functions and more than one hidden layers lead to the so-called deep autoencoders \cite{Brunton2020}. 
\black

Reinforcement learning has aspects of both supervised and unsupervised learning methods. This type of learning refers to a type of interactive learning in which a learning agent evaluates the data space based on its knowledge, and with each step forward an associated reward is assigned to an action taken. The next action is also determined based on this reward. Reinforcement learning models learn by interacting with their environment. The essential aspect of reinforcement learning is determining a policy for how to behave, the reward for setting a goal, and a value function that defines the overall reward in the long run \cite{sutton2018reinforcement,tsantekidis2022DLbookRL}.

\subsection{Reduced-order models}
In reduced-order modeling, the basic idea is to derive a low-dimensional and efficient representation of high-fidelity models \cite{Antoulas01asurvey}. ROMs generally need high-fidelity data as the first input. The development of a ROM consists of two steps: first, finding a set of spatial basis functions (or modes) which define a low-dimensional subspace where the system can be modeled. There are some options to do that including Lagrange basis, Hermite basis, POD basis, Taylor basis, etc. \cite{ito1998reduced}. For data-driven basis functions, a set of simulation data, called snapshots, are used to construct this subspace. The second step of model development is to determine the dynamics of the basis functions’ coefficients which, as for finding the basis functions, can be done in several different ways. The POD technique can be considered one of the most widely used methods to find the basis function because of its capability to deal with the complexity of dynamical systems \cite{bergmann2005optimal}; therefore, the second step in the development of ROM is briefly explained for a POD-based ROM. Traditionally, the coefficients of basis functions can be evaluated by solving ordinary differential equations (ODE) which are derived from the governing equations by using projection methods such as Galerkin projection \cite{hijazi2021podgalerkin}. Since these methods rely on the knowledge of governing equations, they are categorized as PGDDMs and are called intrusive reduced-order models (IROM). There are other methods that build the relationship between the design parameters and mode coefficients without explicit knowledge of the governing equations by following a black-box approach, also called non-intrusive reduced-order models (NIROM) \cite{Chen2021}. In some of the studies using a NIROM approach, ML algorithms play the role to find the coefficients of basis functions (Fig.~\ref{fig:Fig4}b). 

The most important NIROMs can be achieved by using Taylor series expansions \cite{chen2004reduced}, constructing nonlinear ODEs between state variables and input parameters and then solving the problem with the least square method \cite{axelsson1987generalized}, sparse grid method \cite{lin2017non}, coupling POD with nonlinear interpolation methods including radial basis function (RBF) \cite{majdisova2017radial}, Kriging method \cite{xiao2010model}, and ANN, and utilizing dynamic mode decomposition (DMD) method \cite{Schmid2010}. 

Here a brief insight about the DMD technique is provided which is an emerging \blue PDDM \black to obtain ROMs for high-dimensional dynamical systems. The DMD method was first developed by  Schmid in 2008 \cite{Schmid2010}. The method has been applied to a broad range of applications including fluid dynamics, robotics, finance, etc. The DMD can extract from data the spatiotemporal structures that dominate the observed data from that dynamical system. It can also provide a linear dynamical system that shows how they can evolve in time. The first step in a DMD model is collecting data and breaking it into snapshots. As shown in Fig.~\ref{fig:Fig4}c, DMD attempts to find the best linear operator (\textbf{A}) that advances \textbf{X} into $\textbf{X}^{\prime}$. In a complex flow, matrix \textbf{A} has many elements that we do not desire to know. What DMD actually does is approximating the leading eigenvalues and eigenvectors of the \textbf{A} matrix without calculating the matrix itself. $\widetilde{\textbf{A}}$ has the same eigenvalues as the big \textbf{A} matrix, and it is the linear best fit that tells us how POD modes evolve in time. By capturing the eigenvalues of $\widetilde{\textbf{A}}$, one can reach eigenvalues of \textbf{A}. The eigenvalues and eigenvectors of the reduced dynamic operator $\widetilde{\textbf{A}}$ are named as \textbf{W} and $\Lambda$. Once the  eigenvalues and eigenvectors of \textbf{A} are in hand, essentially one can interpret any eigenvector of \textbf{A} has the shape of snapshot column vectors and can be reshaped into the flow field. Therefore, it can be used to predict how the system will evolve in the future \cite{Taira2017}. 

\section{Data-driven wind-farm flow modeling}
\label{section:Results}
In this section, the objective, utilized data, and methodology of the studies focusing on data-driven wind-farm flow modeling are discussed in two groups based on their approach, i.e., \blue PDDM \black and PGDDM approaches.
The majority of studies reviewed in this article \cite{Iungo2015,debnath2017towards,Hamilton2018,Zhang2020,Ali2020,Ali2021,Ali2021b,Chen2022, Wilson2017,Ti2020,Zhang2022,Renganathan2021,Nai-Zhi2022,ZhangZexia2022,Optis2019,Japar2014,Yin2019,Ti2021} follow the \blue PDDM \black approach or the black-box modeling without the need of knowing the governing equations of the problem. But we can see that some recent works from Howland and Dabiri \cite{Howland2019}, Yan et al. \cite{Yan2019}, Park and Park \cite{Park2019}, and Sun et al. \cite{Sun2020} have introduced physics into their data-driven analysis; therefore, they can be labeled as PGDDMs. 
A summary of the papers can be found in Table~\ref{longtable:1}.  
The studies reviewed in this article generally use data from three main sources: 1) wind-tunnel experiments, 2) field measurements using supervisory control and data acquisition (SCADA) and light detection and ranging (LiDAR), and 3) numerical simulations including high-fidelity data from LES, mid-fidelity data from RANS simulations, and low-fidelity data from analytical wake models.
Note that some recent works use a PGDDM approach to predict the flow field before reaching the wind farm (see, e.g., Refs. \cite{ZhangEn2021,ZhangEn2021a}). These works have the same approach and methodology as the papers covered in this review but with a different objective; therefore, they are not covered here.

\subsection{Purely data-driven approach}
\begin{figure}[ht] 
	\centering
	\includegraphics[width=1\textwidth]{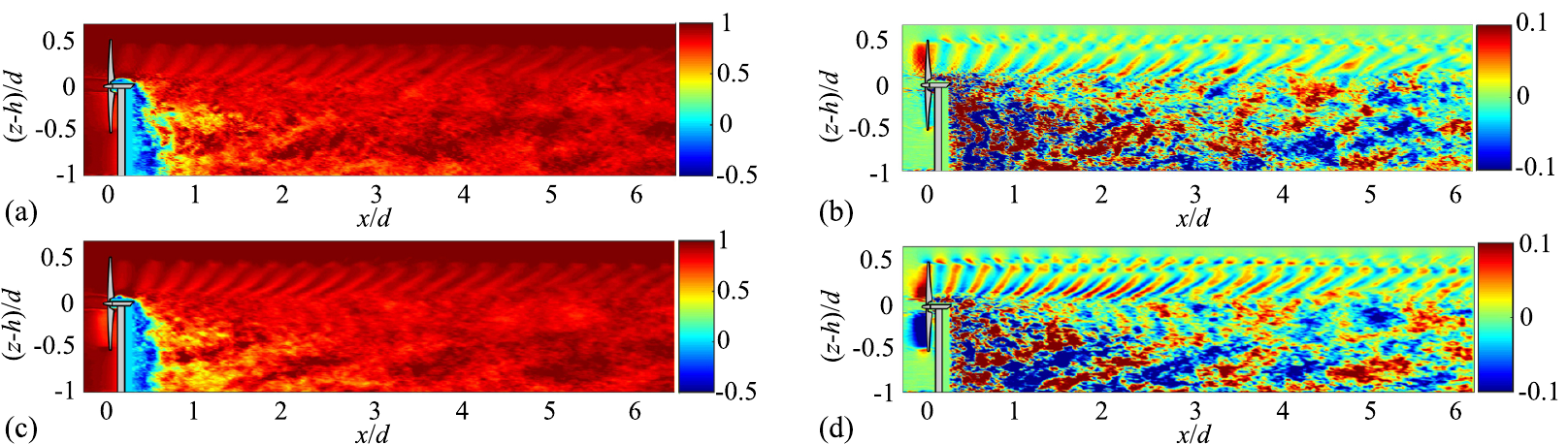}
	\caption{Performance of a ROM developed by using the DMD method: (a) instantaneous streamwise velocity field from LES, \blue (b) streamwise velocity fluctuation field from LES, (c) predicted instantaneous velocity field, and (d) predicted velocity fluctuation field. \black  Here, $d$ is the rotor diameter, $h$ is the hub height, and the velocity is normalized with the incoming wind at the hub height. \blue Reprinted from Iungo, G. V., Santoni-Ortiz, C., Abkar, M., Porté-Agel, F., Rotea, M. A., Leonardi, S., J. Phys.: Conf. Ser., Vol. 625, Article ID012009, 2015; licensed under a Creative Commons Attribution (CC BY) license.} \black
	\label{fig:Fig5}
\end{figure}

Iungo et al. \cite{Iungo2015} generated a ROM to predict the instantaneous flow field behind a wind turbine. They utilized two LES datasets generated at UT Dallas and EPFL. Both datasets used in their study were generated for a neutral ABL. For the first dataset, a model wind turbine with a rotor diameter of 0.89 m was simulated using the actuator-line model \cite{Sorensen2015}, and the mean inflow wind speed at hub height was 10 m/s. By evaluating the axial velocity field at the vertical symmetry-plane of the wake with a sampling frequency of 1124 Hz, 214 snapshots were generated. For generating the second set of LES data, the actuator-disk model with rotation \cite{Abkar2015} was used to model the flow around a Vestas V80-2MW wind turbine with rotor diameter and hub height of 80 m and 70 m, respectively. The mean wind speed at hub height was equal to 8 m/s. For more information on the second dataset, see Ref. \cite{Abkar2015}.
By using the DMD technique, a ROM embedded in the Kalman filter \cite{houtekamer1998data} was developed and validated with the LES data as shown in Fig.~\ref{fig:Fig5}. By comparing the contours of instantaneous streamwise velocity in the middle vertical plane perpendicular to the turbine, it can be concluded that the model can capture the main flow structure and dynamics in the wake region downstream.  

Debnath et al. \cite{debnath2017towards} investigated the coherent vorticity structures within the wake region of a wind turbine with two techniques of POD and DMD. They performed two groups of LESs to generate the data: for the first one, the actuator-line model was used, and for the other one, the turbine's tower and nacelle were also included in the simulation by using the immersed boundary method \cite{orlandi2006dns}. The two sets of LES data held 282 and 387 snapshots, respectively. The turbine had a rotor diameter of 0.89 m, and a uniform  velocity of 10 m/s was used as the inflow condition. The results revealed that, under the laminar inflow condition, by selecting only 6 POD modes, the wake vorticity structures were captured with acceptable accuracy when compared to the LES data. They also analyzed the effect of mode selection on the accuracy of their models.

Wake flow in an array of wind turbines was analyzed in the study of Ali and Cal \cite{Ali2020}.
To generate the data, a parallel experimental study was conducted, and a $3\times3$ array of turbines was placed in the Corrsin wind tunnel at John Hopkins University. Measurements were recorded with a resolution of 40 kHz, resulting in a dataset with 4 million data points. An active grid was placed at the tunnel inlet to introduce turbulence to the flow.
They followed a \blue PDDM \black approach using Hankel-based DMD \cite{arbabi2017ergodic}. 
They performed a short-term forecast of flow-field evolution in the wake of a series of turbines validated with experimental data derived from wind-tunnel tests. The results showed that the mean relative error between the experimental results and prediction was about 15\% for the fluctuating velocities. 

Hamilton et al. \cite{Hamilton2018}, to study flow in a wind farm, after finding the most important modes which could characterize the dynamic wake with a fair accuracy using the POD approach, developed a model with polynomial coefficients which shared the shape with Galerkin-projection-POD \cite{hijazi2021podgalerkin}, but since it uses LES data to find coefficients, it lies in the area of data-driven approaches. 

To identify the coherent structures in the wake region of a single wind-turbine wake, in the study of Ali et al. \cite{Ali2021}, an unsupervised clustering method based on POD was utilized by using LES data. The features of the fluctuating velocity were clustered based on their similarity and were determined as the centroids of the clusters. 
In another study, Ali et al. \cite{Ali2021b} attempted to develop a model by using the k-means clustering method \cite{ahmad2007k} and LES data to make a connection between several linear ROMs and to evaluate the optimum sparse sensor location downstream of a turbine. They aimed to dynamically find the positions which measurements at those points are very critical. As the second stage in their work, they used a long short-term memory (LSTM) network \cite{hochreiter1997long} which was an RNN to predict the fluctuating velocity at the optimal locations of sensors. The results of Refs. \cite{Iungo2015,Hamilton2018,Ali2021,Ali2021b} may also be valid for the flow in a wind farm for the cases where the wake recovery occurs before the flow reaches the downstream turbine.
It should be noted that in the aforementioned studies by Hamilton et al. \cite{Hamilton2018} and Ali et al. \cite{Ali2021, Ali2021b}, the LES code, LESGO \cite{calaf2010large}, was used to generate the data. The turbine, with a hub height and rotor diameter of 100 m, was modeled using the actuator-disk model with rotation. The simulations were performed for a single wind turbine. The mean inflow velocity and turbulence intensity (TI) at hub height were equal to 14.4 m/s and 6\%, respectively. 2000 snapshots were extracted for the analysis with a time resolution of 2 s.

Zhang and Zhao \cite{Zhang2020} developed an ML-based ROM to model dynamic wake in a wind farm. They generated the LES data using the simulator for wind-farm applications (SOWFA) solvers \cite{churchfield2012overview}. The computational domain held three turbines in a row, where the turbines were modeled with the actuator-line technique \cite{Sorensen2015}. Using a time step equal to 0.02 s, the simulation was executed for 1110 s. For the inflow wind speed, three values of 8, 9, and 10 m/s were considered while the TI was 6\%. For each inflow condition, 20 simulations were done by considering yaw angles from -30$^{\circ}$ to 30$^{\circ}$. Then, the flow field at hub height was sampled every second. To be sure that turbine wakes were well developed, the first 400 snapshots from the data were removed, and the dataset was converted into 710 snapshots.
In their study, POD was used for dimension reduction. To predict the low dimensional representation in the next time step, an LSTM network was utilized. After reducing the dimensions with POD, based on the historical behavior of the flow derived from LES data, a supervised ML model was utilized to predict the reduced coefficients at the current time. The inflow velocity, as well as the distributed control parameters from time step 1 to $t$, were introduced as the inputs of the ML algorithm, and the inputs were required to calculate the flow field at the new time of $t+1$. 
\blue
To validate the model, two cases were considered: a single turbine and two turbines in a row. The flow fields predicted by the model were compared to LES data up to 20 s. The authors showed that the developed model had a similar accuracy as LES and similar computational cost as static wake models. 
Then, two other cases were investigated for further time steps. The first case was a single turbine with known yaw change, and the second case was a $3\times3$ turbine array. The predictions for the last two cases were not compared to any reference data. 
\black 

\small
\begin{center}
\begin{longtable}{ l  p{1.5cm} p{4.0cm}  p{3.7cm} p{4.6cm}}
\caption{Summary of papers on data-driven wind-farm flow modeling.}
\label{longtable:1} \\
\hline \multicolumn{1}{l}{Authors} & \multicolumn{1}{l}{Approach} & \multicolumn{1}{l}{Objective} & \multicolumn{1}{l}{Data} & \multicolumn{1}{l}{Method}\\ \hline  
\endfirsthead
\multicolumn{5}{c}%
{{\tablename\ \thetable{} -- continued from previous page}} \\
\hline \multicolumn{1}{l}{Authors} & \multicolumn{1}{l}{Approach} & \multicolumn{1}{l}{Objective} & \multicolumn{1}{l}{Data} & \multicolumn{1}{l}{Method}\\ \hline  
\endhead
\multicolumn{5}{r}{{Continued on next page}} \\ \hline 
\endfoot
\endlastfoot
Iungo et al. \cite{Iungo2015}	&	PDDM	&	Dynamic wake modeling	&	LES	&	DMD	\\
Debnath et al. \cite{debnath2017towards} & PDDM & Dynamic wake modeling & LES & POD, DMD \\
Hamilton et al. \cite{Hamilton2018}	&	PDDM	&	Dynamic wake modeling	&	LES	&	POD-polynomial expansion	\\
Zhang and Zhao \cite{Zhang2020}	&	PDDM	&	Dynamic wake modeling	&	LES	&	POD-RNN	\\
Ali and Cal \cite{Ali2020}	&	PDDM	&	Dynamic wake modeling	&	Wind tunnel experiments	&	DMD	\\
Ali et al. \cite{Ali2021}	&	PDDM	&	Dynamic wake modeling	&	LES	&	POD-k-means	\\
Ali et al. \cite{Ali2021b}	&	PDDM	&	Dynamic wake modeling	&	LES	&	POD-k-means, RNN	\\
Chen et al. \cite{Chen2022}	&	PDDM	&	Dynamic wake modeling	&	LES	&	Extended DMD	\\
Wilson et al. \cite{Wilson2017}	&	PDDM	&	Static wake modeling	&	RANS simulations	&	RF	\\
Ti et al. \cite{Ti2020}	&	PDDM	&	Static wake modeling	&	RANS simulations	&	BPNN	\\
Zhang and Zhao \cite{Zhang2022}	&	PDDM	&	Static wake modeling	&	Time-averaged LES	&	GAN	\\
Renganathan et al. \cite{Renganathan2021}	&	PDDM	&	Static wake modeling	&	LiDAR measurements	&	ANN, GP regression	\\
Nai-zhi et al. \cite{Nai-Zhi2022}	&	PDDM	&	Static wake modeling	&	SCADA	&	GA, RF	\\
Zhang et al. \cite{ZhangZexia2022}	&	PDDM	&	Static wake modeling	&	LES	&	CNN	\\
Optis and Perr-Sauer \cite{Optis2019}	&	PDDM	&	Dynamic power prediction	&	SCADA	&	GAM, ANN, GBM, ERT, SVM	\\
Japar et al.  \cite{Japar2014}	&	PDDM	&	Static power prediction	&	SCADA	&	Linear regression, ANN, SVR	\\
Howland and Dabiri  \cite{Howland2019}	&	PGDDM	&	Static power prediction	&	SCADA	&	ANN	\\
Yan et al. \cite{Yan2019}	&	PGDDM	&	Static power prediction	&	SCADA	&	ANN	\\
Park and Park \cite{Park2019}	&	PGDDM	&	Static power prediction	&	Analytical wake model	&	GNN	\\
Yin and Zhao \cite{Yin2019}	&	PDDM	&	Static power prediction	&	Analytical wake model	&	GRNN, RF, SVM, GBR, RNN	\\
Sun et al. \cite{Sun2020}	&	PGDDM	&	Static power prediction	&	SCADA	&	ANN	\\
Ti et al. \cite{Ti2021}	&	PDDM	&	Static power prediction	&	RANS simulations	&	BPNN	\\

\hline
\end{longtable}
\end{center}
\normalsize

Chen et al. \cite{Chen2022} developed an estimator using Koopman operator theory and extended DMD method \cite{korda2018convergence} to predict the flow in the wake region of a wind turbine. To generate the data, 3D flow in a computational domain with a DTU 10MW turbine \cite{bak2013dtu} was simulated using SOWFA \cite{churchfield2012overview}. The turbine had a hub height and rotor diameter of 119 m and 178.3 m, respectively. The streamwise velocity in the downstream region of the turbine was recorded with a 2 s interval, forming 750 snapshots. Then, a Koopman-linear model was developed to predict the flow in the wake region of the turbine. The developed flow estimator was based on a dynamic state-space model with physical states. To directly provide information from the wake region, probe sensors were considered in the studied space and the locations of sensors were optimized. An analysis of the results of the extended DMD showed a robust performance under with- and without-noise scenarios. 

Several studies with the \blue PDDM \black approach utilize ML tools. 
Wilson et al. \cite{Wilson2017} developed an ML model to predict the static velocity field behind a turbine.
To generate the dataset required for training, validating, and testing the developed model, they conducted RANS simulations using ANSYS Fluent \cite{matsson2021introduction} for a range of inflow velocities from 5.5 m/s to 17.5 m/s and extracted the velocity components in the wake region, resulting in 72,800 examples.
The ML model was based on the random forest (RF) method \cite{biau2016random} with inputs including undisturbed velocity, positions relative to the turbine hub in Cartesian and spherical coordinates (three for each), and nonlinear products of Cartesian components, while the output was the wind-velocity components in the wake region. 

In the study of Zhang and Zhao \cite{Zhang2022}, static wake in a wind farm was analyzed. 
SOWFA \cite{churchfield2012overview} was used to generate the utilized data by simulating flow over three turbines on a row. Then, the flow field around each turbine at the hub-height level was extracted from data and was time-averaged, generating three training samples from each LES simulation. The values for inflow wind speed, TI, and yaw angles considered in this study were similar to the values considered in the previous study of the same authors \cite{Zhang2020}. Finally, 270 training samples, at the expense of 1 million CPU hours, were generated and utilized in their study.
The generative adversarial network (GAN) model \cite{atienza2018advanced} developed in their study had two parts, a generator and a discriminator. The generator took the flow parameters as the input, and after a dense layer and a reshape layer and a series of transposed convolution layers, returned it as the predicted time-averaged flow field. In the intermediate layers, the LeakyReLU activation function was used, and hyperbolic tangent function was used for the last layer. The discriminator part took the flow parameter together with the high-fidelity and the predicted flow fields as the input and finally returned a classification indicator of fake or real. In the intermediate layers, the LeakyReLU activation function was used and the sigmoid function was used for the last layer. The discriminator learned to detect the pair of the real flow field and flow parameters from the pair of the predicted flow field and flow parameter until the generated flow field was not distinguishable from the real flow field. Therefore, the discriminator learned to minimize classification error. In this work, \textit{Keras} package \cite{keras2015theano} with \textit{Tensorflow} \cite{abadi2016tensorflow} were used to develop the model. They used their model to evaluate a wind farm comprised of 6 turbines in two rows under different inflow conditions and yaw angles.

Renganathan et al. \cite{Renganathan2021} used deep autoencoders \cite{atienza2018advanced} to find a low-dimensional latent space that could closely approximate the wake measured with LiDAR with a low computational cost. 
They used LiDAR data recorded during 19 months from a wind farm with 25 identical turbines located in North Texas. For more information on the data-gathering process, see Refs. \cite{el2017quantification,zhan2020lidar}. While the name of the farm was not explicitly mentioned in their paper, information about the turbines was provided. The turbines in the studied wind farm have a rotor diameter of 127 m and a hub height of 89 m, each with a power capacity of 2.3 MW. Meteorological and SCADA data were also gathered with ten-minute averaging for wind speed, wind direction, temperature, pressure, power, the angular speed of the rotor, and blade pitch angle. Mean hub height wind speed and TI recorded by SCADA were in the range of 2.92 m/s to 15.22 m/s and 4\% to 36\%, respectively.
Then, they utilized ANN to map parameter space to mean wake-flow fields (latent space). They also used probabilistic ML with Gaussian process (GP) \cite{seeger2004gaussian} modeling to map parameter space to latent space. To be able to train based on large data, variational GP models \cite{hamelijnck2021spatio} were utilized which performed better than the conventional one. The ANN had an initial layer called the encoder. The rest was the decoder part with the objective to predict flow which was very similar to LiDAR measurements.  

In the study of Nai-zhi et al. \cite{Nai-Zhi2022}, a model was developed to evaluate the wake growth rate from the inflow data since the wake growth rate is an important parameter in most of the analytical wake models. They wanted to make it possible for the analytical wake model to adapt to the incoming flow conditions and improve its accuracy.
SCADA data from two wind farms were utilized in their study. The first farm,  which is anonymous in the paper, is located in Alberta, Canada, with six wind turbines on a single row. All the wind turbines in the first wind farm are Vestas V80, each with a capacity of 2 MW, except one which has a capacity of 1.8 MW. The SCADA data, comprised of 2766 data points, had information on wind speed and wind direction, but not for yaw angle. The second farm is La Haute Borne Wind Farm that is comprised of four Siemens MM82 turbines and is located in north-eastern France. The SCADA data of the second wind farm had ten-minute-recorded information from 2013 to 2016, including wind speed, wind direction, temperature, yaw, and power data.
In the study of Nai-zhi et al. \cite{Nai-Zhi2022}, the base wake model was the Gaussian model \cite{Bastankhah2014}. The model was corrected for the yawed conditions. Wake superposition was also adopted to see the interaction among wakes in a wind farm. The undisturbed wind data including wind speed, wind direction, TI, temperature from SCADA data were used as the input features of the model. First, it calculated the power based on an analytical wake model and solved an optimization problem to minimize the error between measured power and predicted power. The genetic algorithm (GA) \cite{davis1991handbook} was used to find the optimum wake growth rate. Then a machine was trained using the RF method to be able to map the inflow conditions onto the optimal parameters of the wake model. Results showed that by using the model to relate wake growth rate to operational data, an improvement of 20\% compared to analytical models was observed. The results showed that the inflow heterogeneity affects the wake development.

A CNN \cite{albawi2017understanding} autoencoder model was developed by Zhang et al. \cite{ZhangZexia2022} that could predict 3D time-averaged velocity fields at the wake region. 
The data necessary for training, validation, and testing of an ML model were generated by performing LES simulations based on the geometric characteristics of SWiFT facility, Texas (see, e.g., Ref. \cite{berg2014scaled}). This experimental site holds three identical Vestas V27 turbines with rotor diameter and hub height of 27 m and 32.1 m, respectively. For several wind directions, there could be wake losses in the downstream turbine(s). The actuator-surface model \cite{shen2009actuator} was utilized to include the turbines' effect. The LES simulation results were first validated by comparing them against SCADA data and also the field measurements of spinning LiDAR. Four values for the wind direction and five values for the mean inflow velocity (7, 9, 11, 13, and 15 m/s) were considered in the data-generation process. Five instantaneous snapshots with 1000 time-step intervals and time-averaged velocity field comprised each training set in their study.
The CNN autoencoder model had  two parts: an encoder as the first section of the model, with the responsibility to extract features from the data, and a decoder part that outputs the predicted flow field. Investigations were performed to balance the prediction accuracy and the number of learnable parameters. The prediction error of the CNN autoencoder model compared to the time-averaged LES data was less than 3\%, with 88\% less computational costs.

Optis and Perr-Sauer \cite{Optis2019} used nine atmospheric variables as wind speed, wind direction, pressure, temperature, potential temperature gradient, TI, wind shear, turbulent kinetic energy (TKE), and Obukhov length from SCADA data to predict time-series of power generation. They aimed to investigate the influence of atmospheric turbulence and thermal effects on wind-farm power production using a statistical model. 
Hourly data of a wind farm located in Pacific Northwest United States was used in their study. The wind farm is comprised of 100 turbines with a hub height of 80 m. Additional information on the studied wind farm was not provided by the authors due to confidentiality restrictions. The output power from SCADA, recorded for 20 months, was available with a two-minute resolution. Fifteen-minute-averaged meteorological data was also available. After hourly averaging for taking into account the distance between the sensors and wind farm, the dataset held 5,000 observations.
For generalized additive model (GAM) \cite{chouldechova2015generalized} the \textit{Python pyGAM} library, for ANN the \textit{Python}’s \textit{Scikit-learn} \cite{pedregosa2011scikit}, for gradient boosting model (GBM) \cite{bentejac2021comparative} the \textit{Gradient Booster Regressor} module, for extremely randomized trees (ERT) \cite{geurts2006extremely} the \textit{Extra Trees Regressor} module, and for support vector regression (SVR) \cite{awad2015support} the \textit{Support Vector Regressor} module all in \textit{Python}’s \textit{Scikit-learn} library \cite{pedregosa2011scikit} were used. Identification of model hyperparameters was performed through a randomized cross-validation grid search. 
The results showed that TKE was playing a vital role apart from wind speed. They evaluated that TKE is more important than wind direction, pressure, and temperature. They also found out that the type of learning algorithm had a minor effect on the prediction. 

Japar et al. \cite{Japar2014} utilized SCADA data to train and test ML models including linear regression \cite{maulud2020review}, ANN, and SVR to predict the static wake flow in a large wind farm and its power generation. 
In their study, ten-minute-averaged data for the power output of each turbine as well as the wind speed and direction were utilized. The data was extracted from SCADA data from Horns Rev Offshore Wind Farm, Denmark, including 80 Vestas V80-2MW turbines.
Individual-turbine-level power predictions done by using the models showed that predictions made by ANN and SVR were in good agreement with the operational data.

Yin and Zhao \cite{Yin2019} considered mean inflow speed equal to 8, 16, and 28 m/s with a wind direction of 270 deg, and wind directions of 180, 225, and 315 deg with an inflow velocity of 16 m/s, where in all scenarios a TI of 6\% was assumed to generate a database for wind-farm power generation by using flow redirection and induction in steady-state (FLORIS) utility \cite{floris}. 
 They developed a general regression neural network (GRNN) \cite{specht1991general} to predict wind-farm power generation using \textit{Tensorflow} \cite{abadi2016tensorflow} and \textit{NumPy} \cite{oliphant2006guide} of \textit{Python}. The GRNN had three hidden layers with neurons of 512, 256, and 128. A model based on RF was constructed using \textit{RandomForestRegressor} imported from \textit{sklearn.ensemble} in \textit{Scikit-learn} library \cite{pedregosa2011scikit}. SVM was built using SVR borrowed from \textit{sklearn.svm} in \textit{Scikit-learn} library \cite{pedregosa2011scikit}. Gradient boosting regression (GBR) \cite{cai2020prediction} was developed based on an ensemble model from \textit{sklearn} package. RNN was trained by using the \textit{Keras} library \cite{keras2015theano}. The RNN's structure had 50 neurons in the first hidden layer and two output layers for power and thrust. Results of the study showed that RNN and SVM exhibited the best accuracy in predicting the wind-farm power output. RNN was very successful in predicting thrust while GRNN had better computational efficiency.

\begin{figure}[ht] 
	\centering
	\includegraphics[width=0.9\textwidth]{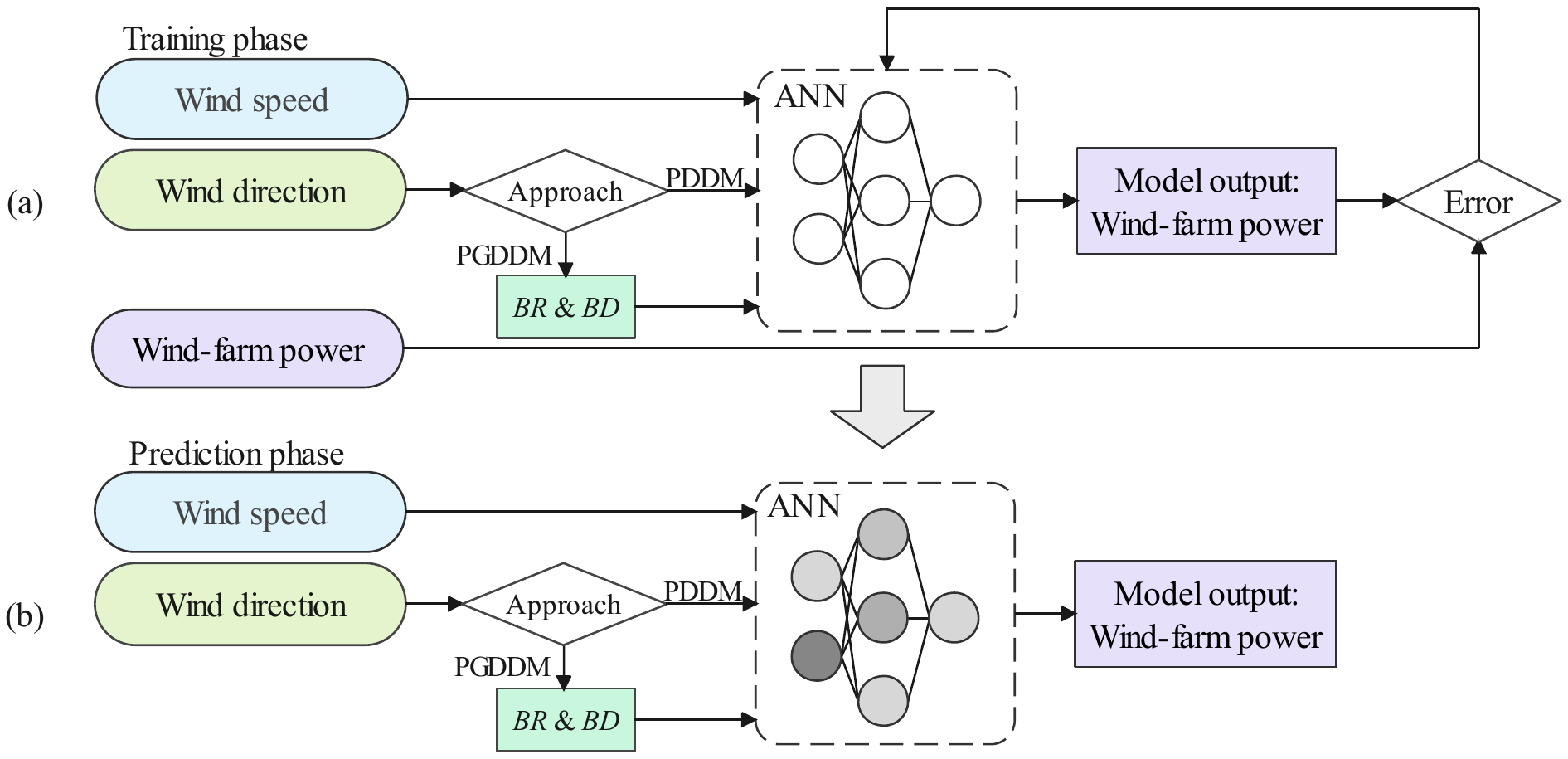}
	\caption{ANN with \blue PDDM \black and PGDDM approaches in the study of Yan et al. \cite{Yan2019}: (a) training phase (b) prediction phase. (Redrawn)}
	\label{fig:Fig6}
\end{figure}

Yan et al. \cite{Yan2019} followed a \blue PDDM \black approach as the first part of their study to predict the power of wind farms.
They utilized data from Lillgrund Offshore Wind Farm, Sweden, which holds 48 Siemens SWT-2.3MW wind turbines. Observation included wind speed, wind directions, yaw angle, and turbine power generation covering 700,000 raw data with a one-minute collection frequency, collected in two and half years. By removing records with yaw bias and the cases when $<90\%$ of turbines were operating, observations were reduced to 204,866 for the analysis. 
The training and prediction phases of their model are shown in Fig.~\ref{fig:Fig6}. They developed an ANN model comprised of two hidden layers with 50 and 40 neurons in each one of them. In the training phase, the inputs were wind direction and wind speed from SCADA data. The power was used to compare the predictions in the backpropagation and update the ANN’s weights in each layer. By using their \blue PDDM\black, they presented power generation and a two-input power curve for Lillgrund Wind Farm, as a function of wind speed and wind direction. Due to the utilized approach, the model developed in this part of their study is not generalizable and may not work with acceptable accuracy for arbitrary wind farms.

The studies of Ti et al. \cite{Ti2020,Ti2021} were based on an ANN model to predict static wake and power in a wind farm. 
They performed RANS simulations for a single Vestas V80-2MW turbine. Results of massive CFD simulations were used to train an ANN that could reproduce velocity and turbulence fields in a wind farm with much less computational cost. For the CFD part, the actuator-disk model was used to consider the turbine effects. The modified $k-\epsilon$ model \cite{mansour1989near} was utilized for turbulence modeling to lower computational costs. A rectangular area that covered the wake of a single turbine was extracted from the data. Then, size of the extracted data was reduced by interpolation. Wind speed and TI in the range of 5-20 m/s and 2\%-26\% were considered in their simulations. 403 simulations were performed to create the dataset.
Wind speed and TI at the hub height were selected as the inputs of the ANN model. To lower the computational cost of ANN, the training data from RANS simulations, which held 24000 elements, were divided into 2000 smaller arrays. Therefore, the ANN model included 2000 sub-models that were trained based on the corresponding data. Every sub-models contained 1 hidden layer with 10 neurons. For hidden and output layers, the tansig and the purelin activation functions were applied. The Levenberg-Marquardt training (trainlm) algorithm \cite{wilamowski2010improved} was used for training which was a fast method. The model was developed in \textit{MATLAB} using \textit{Deep Learning Toolbox} \cite{beale2018deep}. After predicting the wake with the ANN, two popular superposition methods of the sum of square \cite{katic1986simple} and linear \cite{lissaman1979energy} were used to combine the wake effects of several turbines. Then, by using the power formula which was a function of turbine swept area, power coefficient, and wind speed cubed, the power could be found. They applied the ANN model on Horns Rev Offshore Wind Farm and compared its prediction with the operational data, available LES results, and some analytical wake models. 

\begin{figure}[ht]
    \centering
	\includegraphics[width=\textwidth]{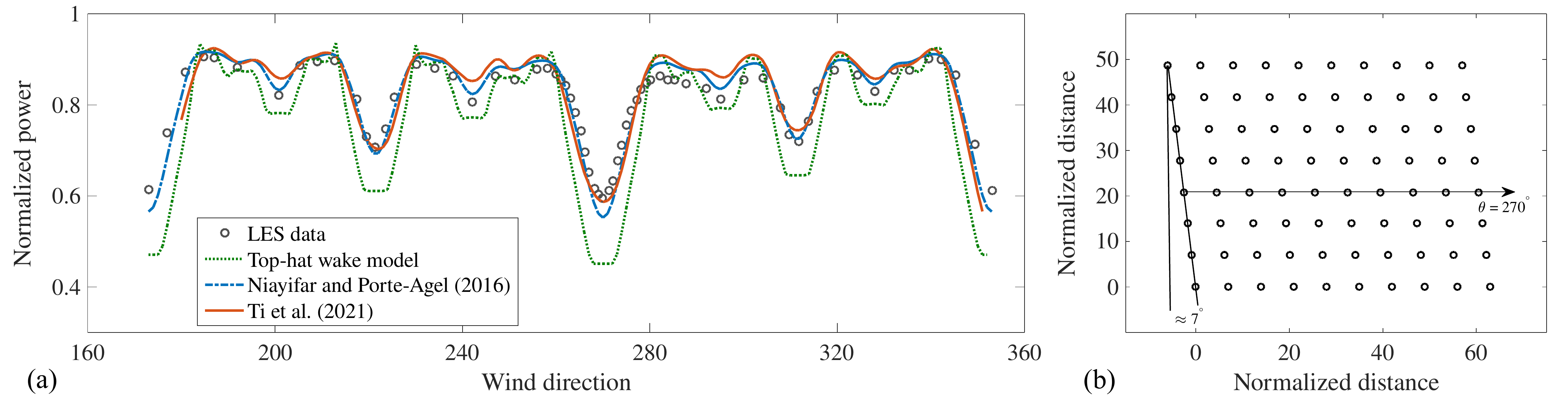}	
    \caption{(a) Comparison of normalized power distribution of Horns Rev Offshore Wind Farm predicted by ANN model of Ti et al. \cite{Ti2021}, analytical-empirical model of Niayifar and Port\'e-Agel \cite{Niayifar2016}, top-hat wake model \cite{katic1986simple}, and LES data of Wu and Port\'e-Agel \cite{Wu2015}. In the top-hat model, the wake growth rate is set to a constant value of 0.04 \cite{Niayifar2016}. (b) The layout of the Horns Rev Wind Farm. Distances are normalized by the rotor diameter.}
    \label{fig:Fig7}
\end{figure}

Here in Fig.~\ref{fig:Fig7}, we have compared the normalized power distribution of Horns Rev Offshore Wind Farm, for wind directions ranging from $173^o - 353^o$, predicted by the ANN model of Ti et al. \cite{Ti2021}, LES \cite{Wu2015}, an analytical-empirical model proposed by Niayifar and Port\'e-Agel \cite{Niayifar2016}, and the extensively utilized top-hat wake model based on the one proposed by Katic et al. \cite{katic1986simple}. The wake model proposed by Niayifar and Port\'e-Agel \cite{Niayifar2016} considers a Gaussian distribution for the velocity deficit, and the evaluation of the local wake growth is performed through local estimation of streamwise TI based on empirical relations. On the other hand, the top-hat wake model assumes a top-hap distribution for the velocity deficit with a constant and spatially uniform wake growth rate within the wind farm. 
Both the ANN model of Ti et al. \cite{Ti2021} and the model of Niayifar and Port\'e-Agel \cite{Niayifar2016} utilize a linear superposition technique to take into account multiple wake interactions, while the top-hat model uses a nonlinear superposition of energy deficits. 
It can be concluded that the ANN model of Ti et al. \cite{Ti2021} has an acceptable accuracy compared to high-fidelity LES data and the wake models. The ANN model has captured the wake velocity distribution and the local wake expansion for the waked turbines by its black-box nature, showing that there is a high potential for the data-driven approach to predict phenomena that cannot be directly modeled by the physics-based models with low computational costs or without the need for empirical relations. A step forward in the study of Ti et al. \cite{Ti2021} could be to use high-fidelity data for the entire wind farm, instead of just one wind turbine, and relax the need of using an empirical superposition method. 

\subsection{Physics-guided data-driven approach}
Turning to studies following a PGDDM approach, in the study of Howland and Dabiri \cite{Howland2019} the idea was that power production of a turbine affected by the wake of upstream wind turbines could be calculated by summation of multiplication of weighing factors and front turbines power production. Therefore, it could be considered a linear regression problem. Having in mind that the thrust coefficient of a turbine is a function of the incoming wind speed, a simple linear relation might not provide a reasonable response. Therefore, the authors proposed some methods: 1) using separate weight values for different wind speed ranges; 2) using an ANN to capture the nonlinearities but the physical interpretation may not be easy; 3) using a nonlinear function for the thrust coefficient as a function of wind speed. 
They utilized one-minute-averaged SCADA data gathered in five years from the Summerview Wind Farm, comprised of 23 Vestas V80-1.8MW wind turbines, located in Alberta, Canada.
A sigmoid function was fitted on the thrust coefficient as a function of generated power. Finally, the power generation of a downstream wind turbine was a summation that needed two learnable parameters. Along with their nonlinear model, they used the linear model and a two-layer ANN to compare the performance of their model. They emphasized the initialization importance, especially in small or shallow ANNs, because ineffective initialization may result in poor performance. Therefore, a physics-informed initialization can help the model to enhance its performance. They used the Gaussian distribution for wakes to initialize the weights. This model required information on the power production of upstream turbines as the only input. Physics has been introduced to their model in two parts as Gaussian profiles: 1) model initialization, and 2) power deficit in model architectures. 

As shown in Fig.~\ref{fig:Fig6}, Yan et al. \cite{Yan2019} also developed an ML model with a PGDDM approach. They used two geometric features of blocking distance (BD) and blockage ratio (BR) as indicators of wake effects on each downstream turbine. By replacing wind direction with BD and BR, the model got more generic and could reduce its site-dependency.
To test the performance of their model, it was applied to the Nørrekær Enge Wind Farm, Denmark, which includes 13 Siemens 2.3MW wind turbines. The SCADA data from the target wind farm was recorded with a ten-minute resolution, holding information on wind speed, yaw angle, and turbines' output power.
Although utilization of the geometric features seems to be very beneficial for the future studies on data-driven wind-farm flow modeling, adding more input features regarding the inflow and wake expansion could be considered and examined to enhance the model performance. Also, a similar data-driven approach can be used to improve the accuracy of wind-farm flow parameterizations in large-scale atmospheric models \cite{fitch2012local,abkar2015new,volker2015explicit,pan2018hybrid}. 

Park and Park \cite{Park2019} presented the wind farm as a graph for any wind condition and used the graph as one of the inputs of their model which was developed to predict the power of the wind farm. They used FLORIS \cite{floris} to calculate the power output of turbines in the generated wind-farm layout. They assumed that the turbines were homogeneous. They also assumed that each turbine was tracking the maximum power point through a control strategy. Then the physics-induced graph neural network (GNN) \cite{scarselli2008graph} received the farm graph and also the power predicted by FLORIS \cite{floris} as the input and predicted the power output of turbines. The model had a weighing function that followed a continuous wake model. The physics-induced weighing function had an exponential function form with some trainable parameters. But since it could result in numerical instability, they approximated it as a five-degree power-series. Their physics-induced GNN had three layers and used the ReLU activation functions. They claimed that their physics-inspired DDM can predict the power generation in any layout and for any wind condition. 

In another study with a PGDDM approach, Sun et al. \cite{Sun2020} developed an ANN model which could predict total static power generation of wind turbines for given wind speed, wind direction, and yaw angle. 
They utilized data from Shiren Wind Farm in northern China to train their model, and five wind turbines were selected to test the model's performance. The UP77-1500 type wind turbines of the selected site have a hub height and rotor diameter of 65 m and 77 m, respectively. The turbines were located on hilly terrain with a maximum altitude difference of 36.7 m. Wind speed, wind direction, yaw angle of turbines, and power output from SCADA data were used for training and testing the machine, while no information on the data resolution is provided.
They used their model to optimize the yaw angles by using a GA. They also attempted to develop a wake model for the wind farm instead of each turbine. Their model was based on the assumption that all turbines were operating.
The features of the ANN included wind speed, wind direction, and yaw angle of wind turbines, while the output was the total power generation. A 2D wake model was used to calculate velocity deficits. The wake model could consider the different heights of turbines. The input layer of the ANN-wake model held wake network (calculated from the wake model), wind speed, and wind direction for five turbines resulting in 15 neurons. The first and second hidden layers had 32 and 64 neurons, respectively, and the total power was the model output. The sigmoid activation function was applied to the first layer and ReLU to the second one. Results showed that informing the wind speed deficit for all turbines to the ANN increased the accuracy of the predictions.

\section{Summary and conclusions}
\label{section:Conclusions}
In this review article, we discussed purely and physics-guided data-driven studies related to wind-farm flow modeling. The reviewed literature was organized based on the modeling approaches, objectives, data, and methodology. This review reveals that the physics-guided data-driven approach has a minor share in data-driven wind-farm flow modeling. Majority of the works utilized machine learning tools to map inputs onto output, which was mostly the farm power output or efficiency. Most of the studies were constructed based on LESs or SCADA data from operational wind farms, but there was no study utilizing data from more than two farms. Moreover, there were some studies completely based on data from analytical wake models. Dynamic wake, as an important feature of the fluid dynamical behavior of wind farms, was not studied in the machine learning-based studies, but the studies with a reduced-order modeling approach, have addressed that phenomenon. Moreover, power generation time-series for wind farms were only investigated and presented in one of the reviewed studies with a \blue purely \black data-driven approach.

An important question that comes to mind when reading articles related to purely and physics-guided data-driven wind-farm flow modeling is whether these models, in a similar way to physics-driven models, can answer a wide range of problems. Is it possible to make an effort to develop more interpretable data-driven models, or are these essentially learning patterns? Answering these questions requires studies with a systematic approach that leaves many doors open for extensive research in the future. A key parameter in a complex problem such as wind-farm flow modeling is the selection of meaningful inputs from weather data, flow, and farm characteristics which needs much more attention and can be used to train more accurate and generalizable models. Moreover, it seems that the focus must shift to physics-guided data-driven models to benefit from their higher accuracy in predictions and higher generalizability compared to purely data-driven models. They also have lower computational cost compared to high-fidelity physics-driven models. 
\blue 
The wind-energy community may also benefit from public trained models. This can allow different research groups to test the performance of the existing data-driven models on various wind farms and effectively attempt to improve them.
\black

Finally, it is worth highlighting the importance of public datasets in data-driven research. 
\blue
The data used in most of the studies reviewed in this paper are not publicly available, or the data generation process is not described at an adequate level to allow other researchers working on the same or similar research problems to reproduce the exact same data for methodology development and experimental comparison.
\black
Taking as examples research fields in which data-driven methodologies achieve top performances, like Computer Vision, Natural Language Processing, and Finance, it can be seen that the introduction of realistic public datasets has led to leaps in performance achieved in the respective research problems, both in terms of benchmarking and in terms of maturity of systems operating in real environments (e.g., commercial systems). Noteworthy examples of such datasets are the ImageNet \cite{deng2009imagenet} and Cifar10 \cite{Krizhevsky09cifar10} image classification datasets, the Large Movie Review Dataset \cite{maas2011learning} for sentiment classification, the KITTI \cite{2012kitti} 3D object detection and tracking dataset, and the FI-2010 \cite{ntakaris2018benchmark} limit order book mid-price direction prediction dataset. 
\blue
We can also highlight an example of a public dataset available for turbulent flows at the Johns Hopkins Turbulence Databases \cite{jhtdb}, which has had a prominent impact on data-driven turbulence research.
\black
This is due to that the availability of such public datasets, accompanied with appropriate experimental protocols, facilitate methodology development, and allows for objective performance evaluation of new methods and fair comparisons with prior methods. 

\section*{Acknowledgment}
The authors acknowledge the financial support from the Independent Research Fund Denmark (DFF) under the Grant No. 0217-00038B. The authors would also like to thank Christoffer Hansen for his comments on the manuscript.

\section*{Conflict of interest}
The authors have no conflicts to disclose.

\section*{Data availability statement}
Data sharing is not applicable to this article as no new data were created in this study.


\bibliographystyle{ieeetr} 

\begin{thebibliography}{}

\end{thebibliography}


\begin{thebibliography}{100}

\bibitem{Potrc2021}
S.~Potr{\v{c}}, L.~{\v{C}}u{\v{c}}ek, M.~Martin, and Z.~Kravanja,
  ``{Sustainable renewable energy supply networks optimization – The gradual
  transition to a renewable energy system within the European Union by 2050},''
  {\em Renewable and Sustainable Energy Reviews}, vol.~146, p.~111186, 2021.

\bibitem{Zappa2019}
W.~Zappa, M.~Junginger, and M.~van~den Broek, ``{Is a 100\% renewable European
  power system feasible by 2050?},'' {\em Applied Energy}, vol.~233-234,
  pp.~1027--1050, 2019.

\bibitem{GLOBALWINDREPORT2021}
``Global wind report 2021,'' tech. rep., Global Wind Energy Council Brussels,
  Belgium, 2021.

\bibitem{veers2019grand}
P.~Veers, K.~Dykes, E.~Lantz, S.~Barth, C.~L. Bottasso, O.~Carlson, A.~Clifton,
  J.~Green, P.~Green, H.~Holttinen, {\em et~al.}, ``Grand challenges in the
  science of wind energy,'' {\em Science}, vol.~366, no.~6464, p.~eaau2027,
  2019.

\bibitem{platis2018first}
A.~Platis, S.~K. Siedersleben, J.~Bange, A.~Lampert, K.~B{\"a}rfuss,
  R.~Hankers, B.~Ca{\~n}adillas, R.~Foreman, J.~Schulz-Stellenfleth, B.~Djath,
  {\em et~al.}, ``First in situ evidence of wakes in the far field behind
  offshore wind farms,'' {\em Scientific Reports}, vol.~8, no.~1, pp.~1--10,
  2018.

\bibitem{lundquist2019costs}
J.~Lundquist, K.~DuVivier, D.~Kaffine, and J.~Tomaszewski, ``Costs and
  consequences of wind turbine wake effects arising from uncoordinated wind
  energy development,'' {\em Nature Energy}, vol.~4, no.~1, pp.~26--34, 2019.

\bibitem{schneemann2020cluster}
J.~Schneemann, A.~Rott, M.~D{\"o}renk{\"a}mper, G.~Steinfeld, and M.~K{\"u}hn,
  ``Cluster wakes impact on a far-distant offshore wind farm's power,'' {\em
  Wind Energy Science}, vol.~5, no.~1, pp.~29--49, 2020.

\bibitem{hansen2006state}
M.~O.~L. Hansen, J.~N. S{\o}rensen, S.~Voutsinas, N.~S{\o}rensen, and H.~A.
  Madsen, ``State of the art in wind turbine aerodynamics and aeroelasticity,''
  {\em Progress in Aerospace Sciences}, vol.~42, no.~4, pp.~285--330, 2006.

\bibitem{hansen2011review}
M.~O. Hansen and H.~Aagaard~Madsen, ``Review paper on wind turbine
  aerodynamics,'' {\em Journal of Fluids Engineering}, vol.~133, no.~11, 2011.

\bibitem{wang2012brief}
T.~Wang, ``A brief review on wind turbine aerodynamics,'' {\em Theoretical and
  Applied Mechanics Letters}, vol.~2, no.~6, p.~062001, 2012.

\bibitem{kheirabadi2019quantitative}
A.~C. Kheirabadi and R.~Nagamune, ``A quantitative review of wind farm control
  with the objective of wind farm power maximization,'' {\em Journal of Wind
  Engineering and Industrial Aerodynamics}, vol.~192, pp.~45--73, 2019.

\bibitem{Shapiro2021}
C.~R. Shapiro, G.~M. Starke, and D.~F. Gayme, ``{Turbulence and control of wind
  farms},'' {\em Annual Review of Control, Robotics, and Autonomous Systems},
  vol.~5, no.~1, 2021.

\bibitem{nash2021wind}
R.~Nash, R.~Nouri, and A.~Vasel-Be-Hagh, ``Wind turbine wake control
  strategies: A review and concept proposal,'' {\em Energy Conversion and
  Management}, vol.~245, p.~114581, 2021.

\bibitem{houck2022review}
D.~R. Houck, ``Review of wake management techniques for wind turbines,'' {\em
  Wind Energy}, vol.~25, no.~2, pp.~195--220, 2022.

\bibitem{Vermeer2003}
L.~J. Vermeer, J.~N. S{\o}rensen, and A.~Crespo, ``{Wind turbine wake
  aerodynamics},'' {\em Progress in Aerospace Sciences}, vol.~39, no.~6-7,
  pp.~467--510, 2003.

\bibitem{Sanderse2011}
B.~Sanderse, S.~P. {Van Der Pijl}, and B.~Koren, ``{Review of computational
  fluid dynamics for wind turbine wake aerodynamics},'' {\em Wind Energy},
  vol.~14, no.~7, pp.~799--819, 2011.

\bibitem{Stevens2017}
R.~J. A.~M. Stevens and C.~Meneveau, ``{Flow structure and turbulence in wind
  farms},'' {\em Annual Review of Fluid Mechanics}, vol.~49, pp.~311--339,
  2017.

\bibitem{PorteAgel2019Review}
F.~Port{\'e}-Agel, M.~Bastankhah, and S.~Shamsoddin, ``{Wind-turbine and
  wind-farm flows: A review},'' {\em Boundary-Layer Meteorology}, vol.~174,
  no.~1, pp.~1--59, 2020.

\bibitem{goccmen2016wind}
T.~G{\"o}{\c{c}}men, P.~Van~der Laan, P.-E. R{\'e}thor{\'e}, A.~P. Diaz, G.~C.
  Larsen, and S.~Ott, ``Wind turbine wake models developed at the {Technical
  University of Denmark}: A review,'' {\em Renewable and Sustainable Energy
  Reviews}, vol.~60, pp.~752--769, 2016.

\bibitem{Archer2018}
C.~L. Archer, A.~Vasel-Be-Hagh, C.~Yan, S.~Wu, Y.~Pan, J.~F. Brodie, and A.~E.
  Maguire, ``{Review and evaluation of wake loss models for wind energy
  applications},'' {\em Applied Energy}, vol.~226, pp.~1187--1207, 2018.

\bibitem{lissaman1979energy}
P.~Lissaman, ``Energy effectiveness of arbitrary arrays of wind turbines,''
  {\em Journal of Energy}, vol.~3, no.~6, pp.~323--328, 1979.

\bibitem{katic1986simple}
I.~Katic, J.~H{\o}jstrup, and N.~O. Jensen, ``A simple model for cluster
  efficiency,'' in {\em European Wind Energy Association Conference and
  Exhibition}, vol.~1, pp.~407--410, A. Raguzzi Rome, Italy, 1986.

\bibitem{voutsinas1990analysis}
S.~Voutsinas, K.~Rados, and A.~Zervos, ``On the analysis of wake effects in
  wind parks,'' {\em Wind Engineering}, pp.~204--219, 1990.

\bibitem{Niayifar2016}
A.~Niayifar and F.~Port{\'e}-Agel, ``{Analytical modeling of wind farms: A new
  approach for power prediction},'' {\em Energies}, vol.~9, p.~741, 2016.

\bibitem{Zong2020}
H.~Zong and F.~Port{\'e}-Agel, ``A momentum-conserving wake superposition
  method for wind farm power prediction,'' {\em Journal of Fluid Mechanics},
  vol.~889, p.~A8, 2020.

\bibitem{Bastankhah2021}
M.~Bastankhah, B.~L. Welch, L.~A. Martínez-Tossas, J.~King, and P.~Fleming,
  ``Analytical solution for the cumulative wake of wind turbines in wind
  farms,'' {\em Journal of Fluid Mechanics}, vol.~911, p.~A53, 2021.

\bibitem{lanzilao2022new}
L.~Lanzilao and J.~Meyers, ``A new wake-merging method for wind-farm power
  prediction in the presence of heterogeneous background velocity fields,''
  {\em Wind Energy}, vol.~25, no.~2, pp.~237--259, 2022.

\bibitem{chowdhury2012unrestricted}
S.~Chowdhury, J.~Zhang, A.~Messac, and L.~Castillo, ``Unrestricted wind farm
  layout optimization ({UWFLO}): Investigating key factors influencing the
  maximum power generation,'' {\em Renewable Energy}, vol.~38, no.~1,
  pp.~16--30, 2012.

\bibitem{shakoor2016wake}
R.~Shakoor, M.~Y. Hassan, A.~Raheem, and Y.-K. Wu, ``Wake effect modeling: A
  review of wind farm layout optimization using {J}ensen's model,'' {\em
  Renewable and Sustainable Energy Reviews}, vol.~58, pp.~1048--1059, 2016.

\bibitem{tao2020nonuniform}
S.~Tao, Q.~Xu, A.~Feij{\'o}o, G.~Zheng, and J.~Zhou, ``Nonuniform wind farm
  layout optimization: A state-of-the-art review,'' {\em Energy}, vol.~209,
  p.~118339, 2020.

\bibitem{Meneveau2019}
C.~Meneveau, ``{Big wind power: Seven questions for turbulence research},''
  {\em Journal of Turbulence}, vol.~20, no.~1, pp.~2--20, 2019.

\bibitem{iungo2015data}
G.~V. Iungo, F.~Viola, U.~Ciri, M.~A. Rotea, and S.~Leonardi, ``Data-driven
  {RANS} for simulations of large wind farms,'' {\em Journal of Physics:
  Conference Series}, vol.~625, no.~1, p.~012025, 2015.

\bibitem{king2018data}
R.~N. King, C.~Adcock, J.~Annoni, and K.~Dykes, ``Data-driven machine learning
  for wind plant flow modeling,'' {\em Journal of Physics: Conference Series},
  vol.~1037, no.~7, p.~072004, 2018.

\bibitem{steiner2021classifying}
J.~Steiner, R.~Dwight, and A.~Vir{\'e}, ``Classifying regions of high model
  error within a data-driven {RANS} closure: Application to wind turbine
  wakes,'' {\em arXiv:2106.15593}, 2021.

\bibitem{Steiner2022}
J.~Steiner, R.~P. Dwight, and A.~Vir{\'{e}}, ``{Data-driven RANS closures for
  wind turbine wakes under neutral conditions},'' {\em Computers \& Fluids},
  vol.~233, p.~105213, 2022.

\bibitem{montans2019data}
F.~J. Mont{\'a}ns, F.~Chinesta, R.~G{\'o}mez-Bombarelli, and J.~N. Kutz,
  ``Data-driven modeling and learning in science and engineering,'' {\em
  Comptes Rendus M{\'e}canique}, vol.~347, no.~11, pp.~845--855, 2019.

\bibitem{Legaard2021}
C.~M. Legaard, T.~Schranz, G.~Schweiger, J.~Drgo{\v{n}}a, B.~Falay, C.~Gomes,
  A.~Iosifidis, M.~Abkar, and P.~G. Larsen, ``Constructing neural network-based
  models for simulating dynamical systems,'' {\em arXiv:2111.01495}, 2021.

\bibitem{Goodfellow2016}
I.~Goodfellow, Y.~Bengio, and A.~Courville, {\em Deep Learning}.
\newblock MIT Press, 2016.

\bibitem{nielsen2015neural}
M.~A. Nielsen, {\em Neural networks and deep learning}, vol.~25.
\newblock Determination Press San Francisco, CA, 2015.

\bibitem{loh2011classification}
W.-Y. Loh, ``Classification and regression trees,'' {\em Wiley
  Interdisciplinary Reviews: Data Mining and Knowledge Discovery}, vol.~1,
  no.~1, pp.~14--23, 2011.

\bibitem{peterson2009k}
L.~E. Peterson, ``K-nearest neighbor,'' {\em Scholarpedia}, vol.~4, no.~2,
  p.~1883, 2009.

\bibitem{fahrmeir2013regression}
L.~Fahrmeir, T.~Kneib, S.~Lang, and B.~Marx, ``Regression models,'' in {\em
  Regression}, pp.~21--72, Springer, 2013.

\bibitem{scholkopf2002learning}
B.~Sch{\"o}lkopf, A.~J. Smola, and F.~Bach, {\em Learning with kernels: support
  vector machines, regularization, optimization, and beyond}.
\newblock MIT press, 2002.

\bibitem{zaras2022DLbook}
A.~Zaras, N.~Passalis, and A.~Tefas, ``Neural networks and backpropagation,''
  {\em {Deep Learning for Robot Perception and Cognition}}, pp.~17--34, 2022.

\bibitem{raitoharju2022DLbook}
J.~Raitoharju, ``Convolutional neural networks,'' {\em {Deep Learning for Robot
  Perception and Cognition}}, pp.~35--69, 2022.

\bibitem{tsantekidis2022DLbook}
A.~Tsantekidis, N.~Passalis, and A.~Tefas, ``Recurrent neural networks,'' {\em
  Deep Learning for Robot Perception and Cognition}, pp.~101--115, 2022.

\bibitem{heidari2022DLbook}
N.~Heidari, L.~Hedegaard, and A.~Iosifidis, ``Graph convolutional networks,''
  {\em {Deep Learning for Robot Perception and Cognition}}, pp.~71--99, 2022.

\bibitem{kiranyaz2017pop}
S.~Kiranyaz, T.~Ince, A.~Iosifidis, and M.~Gabbouj, ``Progressive operational
  perceptrons,'' {\em {Neurocomputing}}, vol.~224, pp.~142--154, 2017.

\bibitem{tran2020heterogeneous}
D.~T. Tran, S.~Kiranyaz, M.~Gabbouj, and A.~Iosifidis, ``Heterogeneous
  multilayer generalized operational perceptron,'' {\em {IEEE Transactions on
  Neural Networks and Learning Systems}}, vol.~31, no.~3, pp.~710--724, 2020.

\bibitem{kiranyaz2020Operational}
S.~Kiranyaz, T.~Ince, A.~Iosifidis, and M.~Gabbouj, ``Operational neural
  networks,'' {\em {Neural Computing and Applications}}, vol.~32,
  pp.~6645--6668, 2020.

\bibitem{kiranyaz2021selfonn}
S.~Kiranyaz, J.~Malik, H.~B. Abdallah, T.~Ince, A.~Iosifidis, and M.~Gabbouj,
  ``Self-organized operational neural networks with generative neurons,'' {\em
  {Neural Networks}}, vol.~140, pp.~294--308, 2021.

\bibitem{jain1999data}
A.~K. Jain, M.~N. Murty, and P.~J. Flynn, ``Data clustering: A review,'' {\em
  ACM Computing Surveys (CSUR)}, vol.~31, no.~3, pp.~264--323, 1999.

\bibitem{jolliffe2016principal}
I.~T. Jolliffe and J.~Cadima, ``Principal component analysis: A review and
  recent developments,'' {\em Philosophical Transactions of the Royal Society
  A: Mathematical, Physical and Engineering Sciences}, vol.~374, no.~2065,
  p.~20150202, 2016.

\bibitem{chatterjee2000introduction}
A.~Chatterjee, ``An introduction to the proper orthogonal decomposition,'' {\em
  Current Science}, pp.~808--817, 2000.

\bibitem{Ringner2008}
M.~Ringn{\'{e}}r, ``{What is principal component analysis?},'' {\em Nature
  Biotechnology}, vol.~26, no.~3, pp.~303--304, 2008.

\bibitem{tenenbaum2000global}
J.~B. Tenenbaum, V.~d. Silva, and J.~C. Langford, ``A global geometric
  framework for nonlinear dimensionality reduction,'' {\em Science}, vol.~290,
  no.~5500, pp.~2319--2323, 2000.

\bibitem{Brunton2020}
S.~L. Brunton, B.~R. Noack, and P.~Koumoutsakos, ``{Machine learning for fluid
  mechanics},'' {\em Annual Review of Fluid Mechanics}, vol.~52, pp.~477--508,
  2020.

\bibitem{sutton2018reinforcement}
R.~S. Sutton and A.~G. Barto, {\em Reinforcement learning: An introduction}.
\newblock MIT Press, 2018.

\bibitem{tsantekidis2022DLbookRL}
A.~Tsantekidis, N.~Passalis, and A.~Tefas, ``Deep reinforcement learning,''
  {\em Deep Learning for Robot Perception and Cognition}, pp.~117--129, 2022.

\bibitem{Antoulas01asurvey}
A.~C. Antoulas, D.~C. Sorensen, and S.~Gugercin, ``A survey of model reduction
  methods for large-scale systems,'' {\em Contemporary Mathematics}, vol.~280,
  pp.~193--219, 2001.

\bibitem{ito1998reduced}
K.~Ito and S.~S. Ravindran, ``A reduced-order method for simulation and control
  of fluid flows,'' {\em Journal of Computational Physics}, vol.~143, no.~2,
  pp.~403--425, 1998.

\bibitem{bergmann2005optimal}
M.~Bergmann, L.~Cordier, and J.-P. Brancher, ``Optimal rotary control of the
  cylinder wake using proper orthogonal decomposition reduced-order model,''
  {\em Physics of Fluids}, vol.~17, no.~9, p.~097101, 2005.

\bibitem{hijazi2021podgalerkin}
S.~Hijazi, M.~Freitag, and N.~Landwehr, ``{POD-Galerkin} reduced order models
  and physics-informed neural networks for solving inverse problems for the
  {Navier-Stokes} equations,'' {\em arXiv:2112.11950}, 2021.

\bibitem{Chen2021}
W.~Chen, Q.~Wang, J.~S. Hesthaven, and C.~Zhang, ``{Physics-informed machine
  learning for reduced-order modeling of nonlinear problems},'' {\em Journal of
  Computational Physics}, vol.~446, p.~110666, 2021.

\bibitem{chen2004reduced}
J.~Chen, S.-M. Kang, J.~Zou, C.~Liu, and J.~E. Schutt-Ain{\'e}, ``Reduced-order
  modeling of weakly nonlinear {MEMS} devices with {Taylor}-series expansion
  and {Arnoldi} approach,'' {\em Journal of Microelectromechanical Systems},
  vol.~13, no.~3, pp.~441--451, 2004.

\bibitem{axelsson1987generalized}
O.~Axelsson, ``A generalized conjugate gradient, least square method,'' {\em
  Numerische Mathematik}, vol.~51, no.~2, pp.~209--227, 1987.

\bibitem{lin2017non}
Z.~Lin, D.~Xiao, F.~Fang, C.~Pain, and I.~M. Navon, ``Non-intrusive reduced
  order modelling with least squares fitting on a sparse grid,'' {\em
  International Journal for Numerical Methods in Fluids}, vol.~83, no.~3,
  pp.~291--306, 2017.

\bibitem{majdisova2017radial}
Z.~Majdisova and V.~Skala, ``Radial basis function approximations: Comparison
  and applications,'' {\em Applied Mathematical Modelling}, vol.~51,
  pp.~728--743, 2017.

\bibitem{xiao2010model}
M.~Xiao, P.~Breitkopf, R.~F. Coelho, C.~Knopf-Lenoir, M.~Sidorkiewicz, and
  P.~Villon, ``Model reduction by cpod and kriging,'' {\em Structural and
  Multidisciplinary Optimization}, vol.~41, no.~4, pp.~555--574, 2010.

\bibitem{Schmid2010}
P.~J. Schmid, ``{Dynamic mode decomposition of numerical and experimental
  data},'' {\em Journal of Fluid Mechanics}, vol.~656, pp.~5--28, 2010.

\bibitem{Taira2017}
K.~Taira, S.~L. Brunton, S.~T. Dawson, C.~W. Rowley, T.~Colonius, B.~J. McKeon,
  O.~T. Schmidt, S.~Gordeyev, V.~Theofilis, and L.~S. Ukeiley, ``{Modal
  analysis of fluid flows: An overview},'' {\em AIAA Journal}, vol.~55, no.~12,
  pp.~4013--4041, 2017.

\bibitem{Iungo2015}
G.~V. Iungo, C.~Santoni-Ortiz, M.~Abkar, F.~Port{\'{e}}-Agel, M.~A. Rotea, and
  S.~Leonardi, ``{Data-driven reduced order model for prediction of wind
  turbine wakes},'' {\em Journal of Physics: Conference Series}, vol.~625,
  no.~1, p.~012009, 2015.

\bibitem{debnath2017towards}
M.~Debnath, C.~Santoni, S.~Leonardi, and G.~V. Iungo, ``Towards reduced order
  modelling for predicting the dynamics of coherent vorticity structures within
  wind turbine wakes,'' {\em Philosophical Transactions of the Royal Society A:
  Mathematical, Physical and Engineering Sciences}, vol.~375, no.~2091,
  p.~20160108, 2017.

\bibitem{Hamilton2018}
N.~Hamilton, B.~Viggiano, M.~Calaf, M.~Tutkun, and R.~B. Cal, ``{A generalized
  framework for reduced-order modeling of a wind turbine wake},'' {\em Wind
  Energy}, vol.~21, no.~6, pp.~373--390, 2018.

\bibitem{Zhang2020}
J.~Zhang and X.~Zhao, ``{A novel dynamic wind farm wake model based on deep
  learning},'' {\em Applied Energy}, vol.~277, p.~115552, 2020.

\bibitem{Ali2020}
N.~Ali and R.~B. Cal, ``{Data-driven modeling of the wake behind a wind turbine
  array},'' {\em Journal of Renewable and Sustainable Energy}, vol.~12, no.~3,
  p.~033304, 2020.

\bibitem{Ali2021}
N.~Ali, M.~Calaf, and R.~B. Cal, ``{Cluster-based probabilistic structure
  dynamical model of wind turbine wake},'' {\em Journal of Turbulence},
  vol.~22, no.~8, pp.~497--516, 2021.

\bibitem{Ali2021b}
N.~Ali, M.~Calaf, and R.~B. Cal, ``{Clustering sparse sensor placement
  identification and deep learning based forecasting for wind turbine wakes},''
  {\em Journal of Renewable and Sustainable Energy}, vol.~13, no.~2, p.~023307,
  2021.

\bibitem{Chen2022}
Z.~Chen, Z.~Lin, X.~Zhai, and J.~Liu, ``{Dynamic wind turbine wake
  reconstruction: A Koopman-linear flow estimator},'' {\em Energy}, vol.~238,
  p.~121723, 2022.

\bibitem{Wilson2017}
B.~Wilson, S.~Wakes, and M.~Mayo, ``Surrogate modeling a computational fluid
  dynamics-based wind turbine wake simulation using machine learning,'' in {\em
  2017 IEEE Symposium Series on Computational Intelligence (SSCI)}, pp.~1--8,
  IEEE, 2017.

\bibitem{Ti2020}
Z.~Ti, X.~W. Deng, and H.~Yang, ``{Wake modeling of wind turbines using machine
  learning},'' {\em Applied Energy}, vol.~257, p.~114025, 2020.

\bibitem{Zhang2022}
J.~Zhang and X.~Zhao, ``{Wind farm wake modeling based on deep convolutional
  conditional generative adversarial network},'' {\em Energy}, vol.~238,
  p.~121747, 2022.

\bibitem{Renganathan2021}
S.~A. Renganathan, R.~Maulik, S.~Letizia, and G.~V. Iungo, ``{Data-driven wind
  turbine wake modeling via probabilistic machine learning},'' {\em Neural
  Computing and Applications}, pp.~1--16, 2022.

\bibitem{Nai-Zhi2022}
G.~Nai-Zhi, Z.~Ming-Ming, and L.~Bo, ``{A data-driven analytical model for wind
  turbine wakes using machine learning method},'' {\em Energy Conversion and
  Management}, vol.~252, p.~115130, 2022.

\bibitem{ZhangZexia2022}
Z.~Zhang, C.~Santoni, T.~Herges, F.~Sotiropoulos, and A.~Khosronejad,
  ``{Time-averaged wind turbine wake flow field prediction using autoencoder
  convolutional neural networks},'' {\em Energies}, vol.~15, no.~1, p.~41,
  2022.

\bibitem{Optis2019}
M.~Optis and J.~Perr-Sauer, ``{The importance of atmospheric turbulence and
  stability in machine-learning models of wind farm power production},'' {\em
  Renewable and Sustainable Energy Reviews}, vol.~112, pp.~27--41, 2019.

\bibitem{Japar2014}
F.~Japar, S.~Mathew, B.~Narayanaswamy, C.~M. Lim, and J.~Hazra, ``{Estimating
  the wake losses in large wind farms: A machine learning approach},'' {\em
  IEEE PES Innovative Smart Grid Technologies Conference (ISGT)}, 2014.

\bibitem{Yin2019}
X.~Yin and X.~Zhao, ``{Big data driven multi-objective predictions for offshore
  wind farm based on machine learning algorithms},'' {\em Energy}, vol.~186,
  p.~115704, 2019.

\bibitem{Ti2021}
Z.~Ti, X.~W. Deng, and M.~Zhang, ``{Artificial neural networks based wake model
  for power prediction of wind farm},'' {\em Renewable Energy}, vol.~172,
  pp.~618--631, 2021.

\bibitem{Howland2019}
M.~F. Howland and J.~O. Dabiri, ``{Wind farm modeling with interpretable
  physics-informed machine learning},'' {\em Energies}, vol.~12, no.~14,
  p.~2716, 2019.

\bibitem{Yan2019}
C.~Yan, Y.~Pan, and C.~L. Archer, ``{A general method to estimate wind farm
  power using artificial neural networks},'' {\em Wind Energy}, vol.~22,
  no.~11, pp.~1421--1432, 2019.

\bibitem{Park2019}
J.~Park and J.~Park, ``{Physics-induced graph neural network: An application to
  wind-farm power estimation},'' {\em Energy}, vol.~187, p.~115883, 2019.

\bibitem{Sun2020}
H.~Sun, C.~Qiu, L.~Lu, X.~Gao, J.~Chen, and H.~Yang, ``{Wind turbine power
  modelling and optimization using artificial neural network with wind field
  experimental data},'' {\em Applied Energy}, vol.~280, p.~115880, 2020.

\bibitem{ZhangEn2021}
J.~Zhang and X.~Zhao, ``{Three-dimensional spatiotemporal wind field
  reconstruction based on physics-informed deep learning},'' {\em Applied
  Energy}, vol.~300, p.~117390, 2021.

\bibitem{ZhangEn2021a}
J.~Zhang and X.~Zhao, ``Spatiotemporal wind field prediction based on
  physics-informed deep learning and {LiDAR} measurements,'' {\em Applied
  Energy}, vol.~288, p.~116641, 2021.

\bibitem{Sorensen2015}
J.~N. S{\o}rensen, R.~F. Mikkelsen, D.~S. Henningson, S.~Ivanell, S.~Sarmast,
  and S.~J. Andersen, ``{Simulation of wind turbine wakes using the actuator
  line technique},'' {\em Philosophical Transactions of the Royal Society A:
  Mathematical, Physical and Engineering Sciences}, vol.~373, no.~2035,
  p.~20140071, 2015.

\bibitem{Abkar2015}
M.~Abkar and F.~Port{\'{e}}-Agel, ``{Influence of atmospheric stability on
  wind-turbine wakes: A large-eddy simulation study},'' {\em Physics of
  Fluids}, vol.~27, no.~3, p.~35104, 2015.

\bibitem{houtekamer1998data}
P.~L. Houtekamer and H.~L. Mitchell, ``Data assimilation using an ensemble
  {Kalman} filter technique,'' {\em Monthly Weather Review}, vol.~126, no.~3,
  pp.~796--811, 1998.

\bibitem{orlandi2006dns}
P.~Orlandi and S.~Leonardi, ``{DNS} of turbulent channel flows with two-and
  three-dimensional roughness,'' {\em Journal of Turbulence}, no.~7, p.~N73,
  2006.

\bibitem{arbabi2017ergodic}
H.~Arbabi and I.~Mezic, ``Ergodic theory, dynamic mode decomposition, and
  computation of spectral properties of the {Koopman} operator,'' {\em SIAM
  Journal on Applied Dynamical Systems}, vol.~16, no.~4, pp.~2096--2126, 2017.

\bibitem{ahmad2007k}
A.~Ahmad and L.~Dey, ``A k-mean clustering algorithm for mixed numeric and
  categorical data,'' {\em Data \& Knowledge Engineering}, vol.~63, no.~2,
  pp.~503--527, 2007.

\bibitem{hochreiter1997long}
S.~Hochreiter and J.~Schmidhuber, ``Long short-term memory,'' {\em Neural
  Computation}, vol.~9, no.~8, pp.~1735--1780, 1997.

\bibitem{calaf2010large}
M.~Calaf, C.~Meneveau, and J.~Meyers, ``Large eddy simulation study of fully
  developed wind-turbine array boundary layers,'' {\em Physics of Fluids},
  vol.~22, no.~1, p.~015110, 2010.

\bibitem{churchfield2012overview}
M.~Churchfield, S.~Lee, and P.~Moriarty, ``Overview of the simulator for wind
  farm application {(SOWFA)},'' {\em National Renewable Energy Laboratory},
  2012.

\bibitem{korda2018convergence}
M.~Korda and I.~Mezi{\'c}, ``On convergence of extended dynamic mode
  decomposition to the {Koopman} operator,'' {\em Journal of Nonlinear
  Science}, vol.~28, no.~2, pp.~687--710, 2018.

\bibitem{bak2013dtu}
C.~Bak, F.~Zahle, R.~Bitsche, T.~Kim, A.~Yde, L.~C. Henriksen, M.~H. Hansen,
  J.~P. A.~A. Blasques, M.~Gaunaa, and A.~Natarajan, ``The {DTU} 10-{MW}
  reference wind turbine,'' in {\em Danish Wind Power Research}, 2013.

\bibitem{matsson2021introduction}
J.~E. Matsson, {\em An introduction to ANSYS Fluent 2021}.
\newblock SDC Publications, 2021.

\bibitem{biau2016random}
G.~Biau and E.~Scornet, ``A random forest guided tour,'' {\em Test}, vol.~25,
  no.~2, pp.~197--227, 2016.

\bibitem{atienza2018advanced}
R.~Atienza, {\em Advanced deep learning with Keras: Apply deep learning
  techniques, autoencoders, GANs, variational autoencoders, deep reinforcement
  learning, policy gradients, and more}.
\newblock Packt Publishing Ltd, 2018.

\bibitem{keras2015theano}
C.~Keras, ``Theano-based deep learning librarycode: https://github.
  com/fchollet,'' {\em Documentation: http://keras. io}, 2015.

\bibitem{abadi2016tensorflow}
M.~Abadi, A.~Agarwal, P.~Barham, E.~Brevdo, Z.~Chen, C.~Citro, G.~S. Corrado,
  A.~Davis, J.~Dean, M.~Devin, {\em et~al.}, ``Tensorflow: Large-scale machine
  learning on heterogeneous distributed systems,'' {\em arXiv:1603.04467},
  2016.

\bibitem{el2017quantification}
S.~El-Asha, L.~Zhan, and G.~V. Iungo, ``Quantification of power losses due to
  wind turbine wake interactions through {SCADA}, meteorological and wind
  {LiDAR} data,'' {\em Wind Energy}, vol.~20, no.~11, pp.~1823--1839, 2017.

\bibitem{zhan2020lidar}
L.~Zhan, S.~Letizia, and G.~Valerio~Iungo, ``{LiDAR} measurements for an
  onshore wind farm: Wake variability for different incoming wind speeds and
  atmospheric stability regimes,'' {\em Wind Energy}, vol.~23, no.~3,
  pp.~501--527, 2020.

\bibitem{seeger2004gaussian}
M.~Seeger, ``Gaussian processes for machine learning,'' {\em International
  Journal of Neural Systems}, vol.~14, no.~02, pp.~69--106, 2004.

\bibitem{hamelijnck2021spatio}
O.~Hamelijnck, W.~Wilkinson, N.~Loppi, A.~Solin, and T.~Damoulas,
  ``Spatio-temporal variational {Gaussian} processes,'' {\em Advances in Neural
  Information Processing Systems}, vol.~34, 2021.

\bibitem{Bastankhah2014}
M.~Bastankhah and F.~Port{\'e}-Agel, ``A new analytical model for wind-turbine
  wakes,'' {\em Renewable Energy}, vol.~70, pp.~116 -- 123, 2014.

\bibitem{davis1991handbook}
L.~Davis, {\em Handbook of genetic algorithms}.
\newblock CumInCAD, 1991.

\bibitem{albawi2017understanding}
S.~Albawi, T.~A. Mohammed, and S.~Al-Zawi, ``Understanding of a convolutional
  neural network,'' in {\em International Conference on Engineering and
  Technology (ICET)}, pp.~1--6, IEEE, 2017.

\bibitem{berg2014scaled}
J.~Berg, J.~Bryant, B.~LeBlanc, D.~C. Maniaci, B.~Naughton, J.~A. Paquette,
  B.~R. Resor, J.~White, and D.~Kroeker, ``Scaled wind farm technology facility
  overview,'' in {\em 32nd ASME Wind Energy Symposium}, p.~1088, 2014.

\bibitem{shen2009actuator}
W.~Z. Shen, J.~H. Zhang, and J.~N. S{\o}rensen, ``The actuator surface model: a
  new {Navier--Stokes} based model for rotor computations,'' {\em Journal of
  Solar Energy Engineering}, vol.~131, no.~1, 2009.

\bibitem{chouldechova2015generalized}
A.~Chouldechova and T.~Hastie, ``Generalized additive model selection,'' {\em
  arXiv:1506.03850}, 2015.

\bibitem{pedregosa2011scikit}
F.~Pedregosa, G.~Varoquaux, A.~Gramfort, V.~Michel, B.~Thirion, O.~Grisel,
  M.~Blondel, P.~Prettenhofer, R.~Weiss, V.~Dubourg, {\em et~al.},
  ``Scikit-learn: Machine learning in python,'' {\em Journal of Machine
  Learning Research}, vol.~12, pp.~2825--2830, 2011.

\bibitem{bentejac2021comparative}
C.~Bent{\'e}jac, A.~Cs{\"o}rg{\H{o}}, and G.~Mart{\'\i}nez-Mu{\~n}oz, ``A
  comparative analysis of gradient boosting algorithms,'' {\em Artificial
  Intelligence Review}, vol.~54, no.~3, pp.~1937--1967, 2021.

\bibitem{geurts2006extremely}
P.~Geurts, D.~Ernst, and L.~Wehenkel, ``Extremely randomized trees,'' {\em
  Machine Learning}, vol.~63, no.~1, pp.~3--42, 2006.

\bibitem{awad2015support}
M.~Awad and R.~Khanna, ``Support vector regression,'' in {\em Efficient
  Learning Machines}, pp.~67--80, Springer, 2015.

\bibitem{maulud2020review}
D.~Maulud and A.~M. Abdulazeez, ``A review on linear regression comprehensive
  in machine learning,'' {\em Journal of Applied Science and Technology
  Trends}, vol.~1, no.~4, pp.~140--147, 2020.

\bibitem{floris}
M.~Sinner, E.~Simley, J.~King, P.~Fleming, and L.~Y. Pao, ``Power increases
  using wind direction spatial filtering for wind farm control: Evaluation
  using {FLORIS}, modified for dynamic settings,'' {\em Journal of Renewable
  and Sustainable Energy}, vol.~13, no.~2, p.~023310, 2021.

\bibitem{specht1991general}
D.~F. Specht, ``A general regression neural network,'' {\em IEEE Transactions
  on Neural Networks}, vol.~2, no.~6, pp.~568--576, 1991.

\bibitem{oliphant2006guide}
T.~E. Oliphant, {\em A guide to NumPy}, vol.~1.
\newblock Trelgol Publishing USA, 2006.

\bibitem{cai2020prediction}
J.~Cai, K.~Xu, Y.~Zhu, F.~Hu, and L.~Li, ``Prediction and analysis of net
  ecosystem carbon exchange based on gradient boosting regression and random
  forest,'' {\em Applied Energy}, vol.~262, p.~114566, 2020.

\bibitem{mansour1989near}
N.~Mansour, J.~Kim, and P.~Moin, ``Near-wall k-epsilon turbulence modeling,''
  {\em AIAA Journal}, vol.~27, no.~8, pp.~1068--1073, 1989.

\bibitem{wilamowski2010improved}
B.~M. Wilamowski and H.~Yu, ``Improved computation for {Levenberg--Marquardt}
  training,'' {\em IEEE Transactions on Neural Networks}, vol.~21, no.~6,
  pp.~930--937, 2010.

\bibitem{beale2018deep}
M.~H. Beale, M.~T. Hagan, and H.~B. Demuth, ``Deep learning toolbox™
  reference,'' {\em The MathWorks, Natick, MA, USA}, 2018.

\bibitem{Wu2015}
Y.~T. Wu and F.~Port{\'{e}}-Agel, ``{Modeling turbine wakes and power losses
  within a wind farm using LES: An application to the Horns Rev offshore wind
  farm},'' {\em Renewable Energy}, vol.~75, pp.~945--955, 2015.

\bibitem{fitch2012local}
A.~C. Fitch, J.~B. Olson, J.~K. Lundquist, J.~Dudhia, A.~K. Gupta,
  J.~Michalakes, and I.~Barstad, ``Local and mesoscale impacts of wind farms as
  parameterized in a mesoscale {NWP} model,'' {\em Monthly Weather Review},
  vol.~140, no.~9, pp.~3017--3038, 2012.

\bibitem{abkar2015new}
M.~Abkar and F.~Port{\'e}-Agel, ``A new wind-farm parameterization for
  large-scale atmospheric models,'' {\em Journal of Renewable and Sustainable
  Energy}, vol.~7, no.~1, p.~013121, 2015.

\bibitem{volker2015explicit}
P.~Volker, J.~Badger, A.~N. Hahmann, and S.~Ott, ``The explicit wake
  parametrisation v1.0: A wind farm parametrisation in the mesoscale model
  {WRF},'' {\em Geoscientific Model Development}, vol.~8, no.~11,
  pp.~3715--3731, 2015.

\bibitem{pan2018hybrid}
Y.~Pan and C.~L. Archer, ``A hybrid wind-farm parametrization for mesoscale and
  climate models,'' {\em Boundary-Layer Meteorology}, vol.~168, no.~3,
  pp.~469--495, 2018.

\bibitem{scarselli2008graph}
F.~Scarselli, M.~Gori, A.~C. Tsoi, M.~Hagenbuchner, and G.~Monfardini, ``The
  graph neural network model,'' {\em IEEE Transactions on Neural Networks},
  vol.~20, no.~1, pp.~61--80, 2008.

\bibitem{deng2009imagenet}
J.~Deng, W.~Dong, R.~Socher, L.-J. Li, K.~Li, and L.~Fei-Fei, ``Imagenet: A
  large-scale hierarchical image database,'' in {\em IEEE Conference on
  Computer Vision and Pattern Recognition}, pp.~248--255, 2009.

\bibitem{Krizhevsky09cifar10}
A.~Krizhevsky and G.~Hinton, ``Learning multiple layers of features from tiny
  images,'' 2009.

\bibitem{maas2011learning}
A.~L. Maas, R.~E. Daly, T.~Pham, Peter, D.~Huang, N.~A. Ng, and C.~Potts,
  ``{Learning word vectors for sentiment analysis},'' {\em The 49th Annual
  Meeting of the Association for Computational Linguistics}, 2011.

\bibitem{2012kitti}
A.~Geiger, P.~Lenz, and R.~Urtasun, ``{Are we ready for autonomous driving? The
  {KITTI} vision benchmark suite},'' in {\em {IEEE} Conference on Computer
  Vision and Pattern Recognition}, pp.~3354--3361, 2012.

\bibitem{ntakaris2018benchmark}
A.~Ntakaris, M.~Magris, J.~Kanniainen, M.~Gabbouj, and A.~Iosifidis,
  ``Benchmark dataset for mid-price prediction of limit order book data,'' {\em
  {Journal of Forecasting}}, vol.~37, pp.~852--866, 2018.

\bibitem{jhtdb}
K.~Kanov, R.~Burns, C.~Lalescu, and G.~Eyink, ``The {Johns Hopkins Turbulence
  Databases}: An open simulation laboratory for turbulence research,'' {\em
  Computing in Science \& Engineering}, vol.~17, no.~5, pp.~10--17, 2015.

\end{thebibliography}





\end{document}